\documentclass[twocolumn]{revtex4-1}
\usepackage[T1]{fontenc}
\usepackage[latin9]{inputenc}
\usepackage{color}
\usepackage{amsmath}
\usepackage{graphicx}
\usepackage[english]{babel}
\begin{document}

\title{Acceleration in Non-planar Time-dependent Billiard}

\author{Sedighe Raeisi}

\affiliation{Department of Physics, Faculty of Sciences, Ferdowsi University of
Mashhad, Mashhad, Iran}

\author{Parvin Eslami}

\affiliation{Department of Physics, Faculty of Sciences, Ferdowsi University of
Mashhad, Mashhad, Iran}
\begin{abstract}
We study the dynamical properties of a particle in a non-planar square
billiard. The plane of the billiard has a sinusoidal shape. We consider
both the static and time-dependent plane. We study the affect of different
parameters that control the geometry of the billiard in this model.
We consider variations of different parameters of the model and describe
how the particle trajectory is affected by these parameters. We also
investigate the dynamical behavior of the system in the static condition
using its reduced phase plot and show that the dynamics of the particle
inside the billiard may be regular, mixed or chaotic. Finally, the
problem of the particle energy growth is studied in the billiard with
the time-dependent plane. We show that when in the static case, the
billiard is chaotic, then the particle energy in the time-dependent
billiard grows\textcolor{red}{{} }for small number of collisions, and
then it starts to saturate. But when the dynamics of the static case
is regular, then the particle average energy in the time-dependent
situation stays constant.
\end{abstract}



\maketitle

\section{Introduction }

Billiard is a dynamical system of a mass point particle which moves
along geodesic lines inside a closed region. When the particle reaches
the boundary, it would reflect elastically. 

Despite the simplicity, billiard has a rich physics \cite{Robnik1983},
which makes it a powerful tool for modeling a vast range of physical
phenomena and systems, from microwave field in resonators \cite{Richter1999}
to semiconductors \cite{Micolich2000}, optics \cite{Milner2001,kaplan2005atom}
and acoustic \cite{Kawabe2003}. 

The dynamics of a billiard is governed by the shape of the boundary
and the geometry of the plane. The static non-planar billiards have
been studied before and it was shown that although the square flat
billiard is integrable, the non-planar square billiard can exhibit
chaos \cite{salazar2012constrained}. Until recently, only billiards
with static curved plane has been studied. Here we investigate the
billiard with time-dependent non-planar surface.

In the time-dependent situation, the particle may accelerate. This
behavior that particles could be accelerated by time-dependent perturbations
of the boundary, is known as Fermi acceleration. In \cite{fermi1949origin}
Fermi explained the origin of heavy ion acceleration of cosmic rays.
Fermi showed that, a classical particle may gain unbounded energy
upon collisions with a heavy and moving wall. This is of particular
interest, especially for understanding the unbounded ( indeterminate)
energy growth in systems that a particle experiences collisions with
a time-dependent boundary. Many billiards with time-dependent boundary
as a two dimensional model of such systems have been studied \cite{leonel2009fermi,loskutov2002particle,Oliveira2010,Carvalho2006,gelfreich2011robust}.
Attempts to explain the relation between exhibition of the Fermi acceleration
in a time-dependent billiard and the nonlinear dynamics of its static
version leaded to the Loskutov-Ryabov-Akinshin (LRA) conjecture. The
LRA conjecture states that the ``chaotic dynamics of a billiard with
a fixed boundary is a sufficient condition for the Fermi acceleration
in the system when a boundary perturbation is introduced'' \cite{Loskutov2000}.
It was cleared that the existence of orbits of unbounded and rapid
energy growth is a general phenomenon, typical for arbitrary slow
non-autonomous perturbation of a Hamiltonian system with chaotic behavior
\cite{Gelfreich2008}. It was also discussed in \cite{gelfreich2008fermi}
that another mechanism can accelerates the particle to unbounded energy.
This showed that the presence of unbounded energy growth is possible
for billiards which the phase space of their frozen case retains the
small horseshoe for all times. The unbounded energy growth also can
be observed in systems with zero Lyapanov exponent\cite{shah2010exponential}. 

The quest for observation of unbounded energy growth and verifying
the validity of the LRA conjecture are two of the motivations for
studying time-dependent billiards in the recent decade \cite{gelfreich2011robust,Carvalho2006,Leonel2004,Leonel2010,Livorati2008,Oliveira2010,Oliveira2010a,oliveira2012scaling,batistic2014exponential}.
The Fermi acceleration has been mostly studied in planar billiards
with time-dependent boundary. However, energy growth due to the time-dependence
of the plane of a billiard has not been studied yet. Here we investigate
this problem and show that time-dependence of the curvature of the
plane of a billiard may lead to some growth in average energy at first
and then the average energy starts to saturate.

These studies could have potential applications in variety of phenomenon
and fields, e.g., for understanding the dynamics of an electron confined
in a curved graphene sheet or a crystal lattice. The graphene can
be deformed to form a curved thin sheet. This thin layer of graphene
can be considered as a plane of a non-planar billiard. This thin layer
has some vibrations which makes its curvature time-dependent. For
some applications like sensitive mass detection and high precision
metrology, it is crucial to take these vibrations into account \cite{dai2012nonlinear}.
Since, the random shaped vibrations of graphene sheets can be expanded
in terms of sinusoidal functions, the time-dependent sinusoidal shaped
geometry is one of the ways to study this system. 

Another application of our work is for cold atoms in optical traps
where a cold atom is confined to a node of a standing wave with large
wave length, thus the atom is approximately restricted to move on
a flat two dimensional space \cite{Milner2001,kaplan2005atom}.

The applications of the investigation of particle behavior in a bounded
time-dependent space can also be extended to general relativity. Recently,
experiments are performed which emulates the general relativity phenomenon,
like the study of the behavior of light in a curved space. This includes
the study and simulation of the behavior of light, wave packets or
solitons in a curved space \cite{Schultheiss2010,Bekenstein2014,Bekenstein2015,Peschel2007}.
These studies open a new horizon in understanding the general relativity
and behaviors of celestial objects and probably new aspects of this
theory.

Here, we study the dynamics of a particle in a rectangular non-planar
billiard, where the plane of the billiard changes as a traveling sinusoidal
wave. When, either of the amplitude or the wave vector of the surface
vanishes, the system would be the well known flat rectangular billiard.
Although the planar rectangular billiard is quiet integrable, we show
that the static case, can be chaotic. In fact, the rectangle is chosen
as the boundary, to show the transition from integrable to chaotic
dynamics in the static situation. This generalizes previous studies
in two aspects. First, we consider a particle in a non-planar billiard
with time-dependent plane curvature. Second, we study the behavior
of the energy growth of a particle in a time-dependent non-planar
billiard.

The outline of our paper is as follows. In section II the theory and
the numerical method are described. In section III the behavior of
the particle is explored numerically. In the Sec. III A, we explain
the effect of parameters of the system on the shape of a trajectory
of the particle. In the Sec. III B, the dynamics of the particle in
the static situation is discussed using the phase space Poincare section
(where the static situation refers to when the wave velocity is zero).
In Sec. III C. we investigate the behavior of the energy growth of
the particle and the application of LRA conjecture to our model. 

Finally, the section IV is the conclusion.

\section{Theory }

In this model, a classical particle is constrained to move on a two
dimensional time-dependent surface which is confined to a contour.
This surface is embedded in the three-dimensional Euclidean space. 

The surface $\varSigma$ is defined as
\begin{equation}
\varSigma:y^{i}=y^{i}\left(u^{1},u^{2};t\right)\mbox{ \ensuremath{}\ensuremath{}\ensuremath{}\ensuremath{}\ensuremath{\mbox{ \ensuremath{}\ensuremath{}\ensuremath{}}}},\mbox{ \ensuremath{}\ensuremath{}\ensuremath{}\ensuremath{}}i=1,2,3,\label{eq:2-1}
\end{equation}
where $y^{i}$'s are Cartesian coordinates in the flat three-dimensional
space and $u{}^{1},\,u^{2}$ are coordinates defined on the two-dimensional
curved surface. The selection of $u{}^{1},\,u^{2}$ depends on surface
symmetries.

The position of the particle constrained to the $\Sigma$, can be
expressed as 
\begin{eqnarray}
\overset{\rightarrow}{r} & = & \overset{\rightarrow}{r}\left(u^{1},u^{2};t\right)=y^{1}\left(u^{1},u^{2};t\right)\hat{i}\nonumber \\
 & + & y^{2}\left(u^{1},u^{2};t\right)\hat{j}+y^{3}\left(u^{1},u^{2};t\right)\hat{k}.\label{eq:2-2}
\end{eqnarray}
The motion of a particle which is constrained to the surface $\Sigma$
will depend on the surface geometry which is specified by the metric
\begin{equation}
a_{\alpha\beta}=\frac{\text{\ensuremath{\partial}}y^{i}(u^{1},u^{2};t)}{\text{\ensuremath{\partial}}u^{\alpha}}\frac{\text{\ensuremath{\partial}}y^{i}(u^{1},u^{2};t)}{\text{\ensuremath{\partial}}u^{\beta}}\mbox{ \ensuremath{}\ensuremath{}\ensuremath{}\ensuremath{}}\alpha\mbox{,}\beta=1,2.\label{eq:2-3}
\end{equation}
The Lagrangian of a particle of mass $\mu$ with a potential $V\left(\overrightarrow{r}\right)$
constrained to the surface $\Sigma$ is
\begin{equation}
L=\frac{\mu}{2}a_{\alpha\beta}(u^{1},u^{2};t)\dot{u}^{\alpha}\dot{u}^{\beta}-V\left(\overrightarrow{r}\right)\mbox{ \ensuremath{}\ensuremath{}\ensuremath{}\ensuremath{}\ensuremath{\mbox{ \ensuremath{}\ensuremath{}\ensuremath{}\ensuremath{}}}},\mbox{ \ensuremath{}\ensuremath{}\ensuremath{}\ensuremath{}}\alpha\mbox{,}\beta=1,2,\label{eq:2-4}
\end{equation}
For simplicity, we consider $V\left(\overrightarrow{r}\right)=0$
and $\mu=1$ in our model.

Using the Eq. \ref{eq:2-4}, canonical momentums and the Euler-Lagrange
equations are
\begin{eqnarray}
p_{\alpha}=\text{\ensuremath{\partial}}_{\dot{u}^{\alpha}}L(u^{1},u^{2};t) & = & a_{\alpha\beta}(u^{1},u^{2};t)\dot{u}^{\beta},\label{eq:2-5}\\
\frac{d}{dt}\left(\frac{\text{\ensuremath{\partial}}L(u^{1},u^{2};t)}{\text{\ensuremath{\partial}}\dot{u}^{\alpha}}\right) & - & \frac{\text{\ensuremath{\partial}}L(u^{1},u^{2};t)}{\text{\ensuremath{\partial}}u^{\alpha}}=0.\label{eq:2-6}
\end{eqnarray}

In our model, we consider the motion of a particle on a traveling
wave surface with a square boundary which, may be aligned in a direction
other than the wave vector. The surface $\Sigma$ is defined as 
\begin{equation}
\left\{ \begin{array}{c}
u^{1}=x\\
u^{2}=y
\end{array}\right.\Longrightarrow\left\{ \begin{array}{c}
y^{1}=u^{1}\\
y^{2}=u^{2}\\
y^{3}=f\left(u^{1},u^{2};t\right)=\\
A\sin\left(k_{x}x+k_{y}y-\omega t\right)
\end{array}\right.,\label{eq:2-7}
\end{equation}
where $A$ and $\omega$ are the amplitude and angular frequency.
The $k_{x}$ and $k_{y}$ are components of the wave vector, $k$.
\begin{eqnarray}
k_{x} & = & k\cos\left(\theta_{wave}\right),\nonumber \\
k_{y} & = & k\sin\left(\theta_{wave}\right).\label{eq:2-8}
\end{eqnarray}
where $k$ and $\theta_{wave}$ are respectively the magnitude of
the wave vector and the angle between the wave vector, $\overrightarrow{k}$,
and the $x$ axis. The magnitude of the wave velocity is specified
by $V_{wave}=\frac{\omega}{k}$.

In this paper, we use \textquotedbl{}wave\textquotedbl{} to refer
to the geometry of the surface.

Figure \ref{fig. 1} gives four snapshots of the motion of the particle
on the time dependent surface. In this figure, the wave fronts are
aligned along the $y$ axis. 

\begin{center}
\begin{figure}
\begin{centering}
\includegraphics[width=0.5\columnwidth]{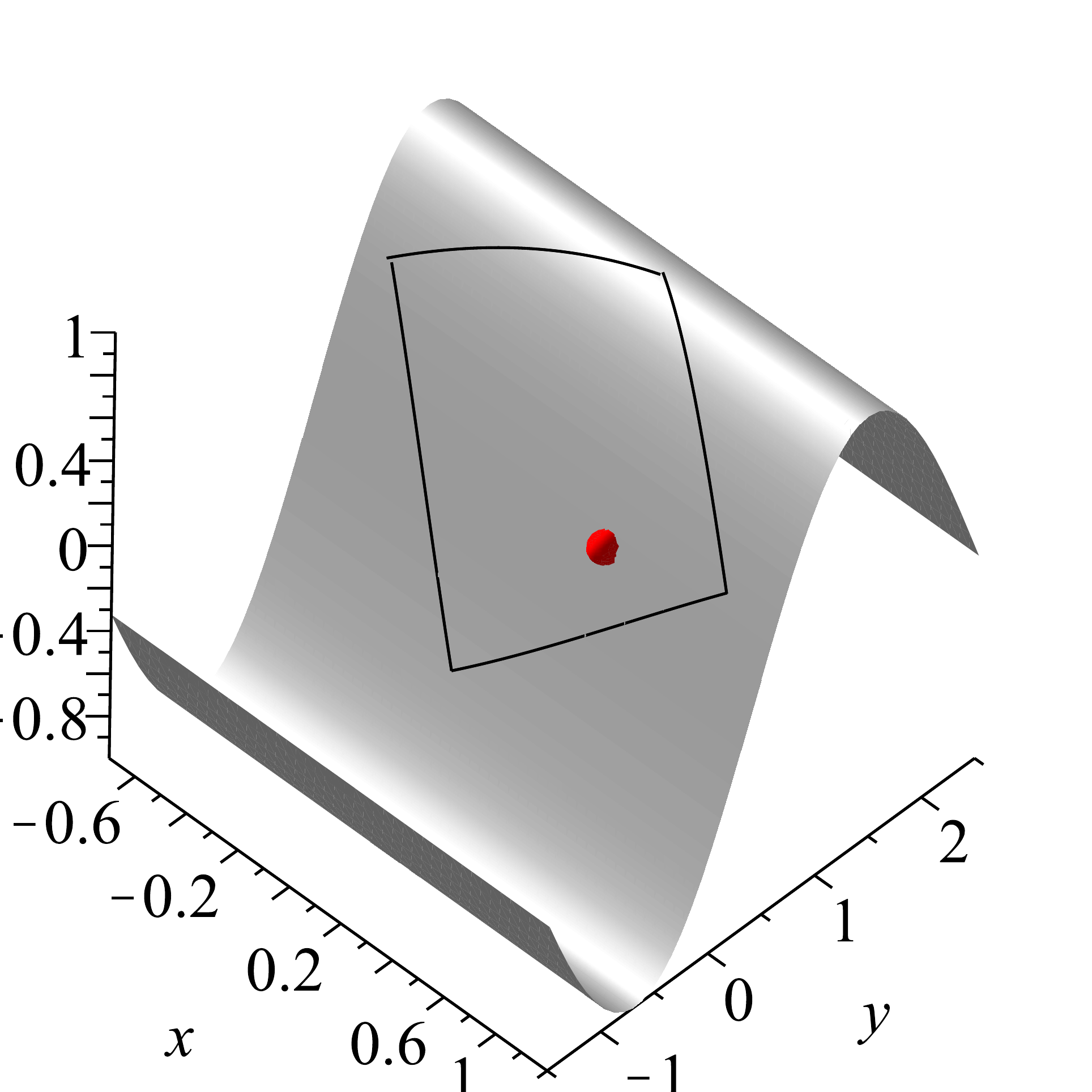}\includegraphics[width=0.5\columnwidth]{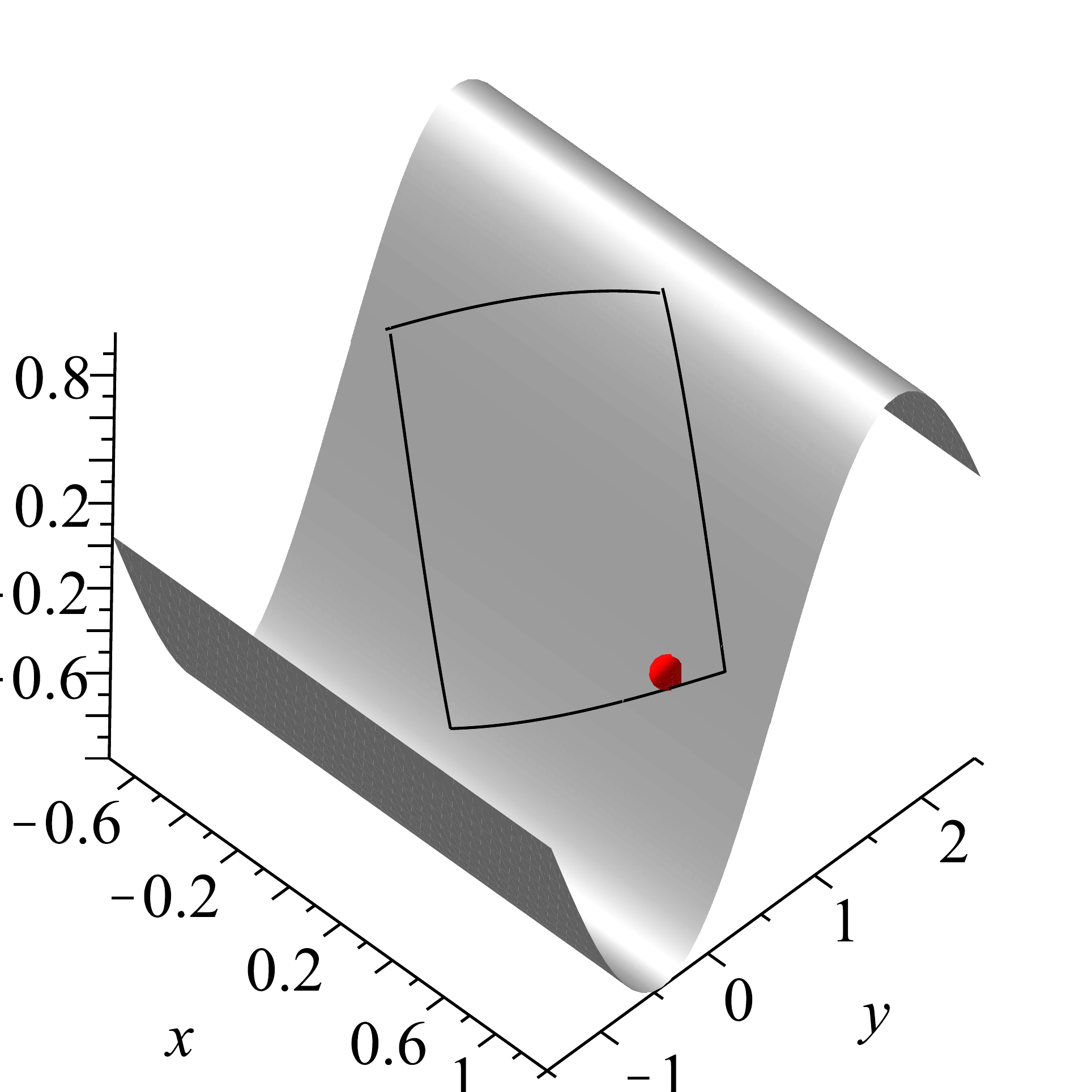}
\par\end{centering}

\begin{minipage}[t]{0.5\columnwidth}%
\begin{center}
(a)
\par\end{center}%
\end{minipage}%
\begin{minipage}[t]{0.5\columnwidth}%
\begin{center}
(b)
\par\end{center}%
\end{minipage}

\begin{centering}
\includegraphics[width=0.5\columnwidth]{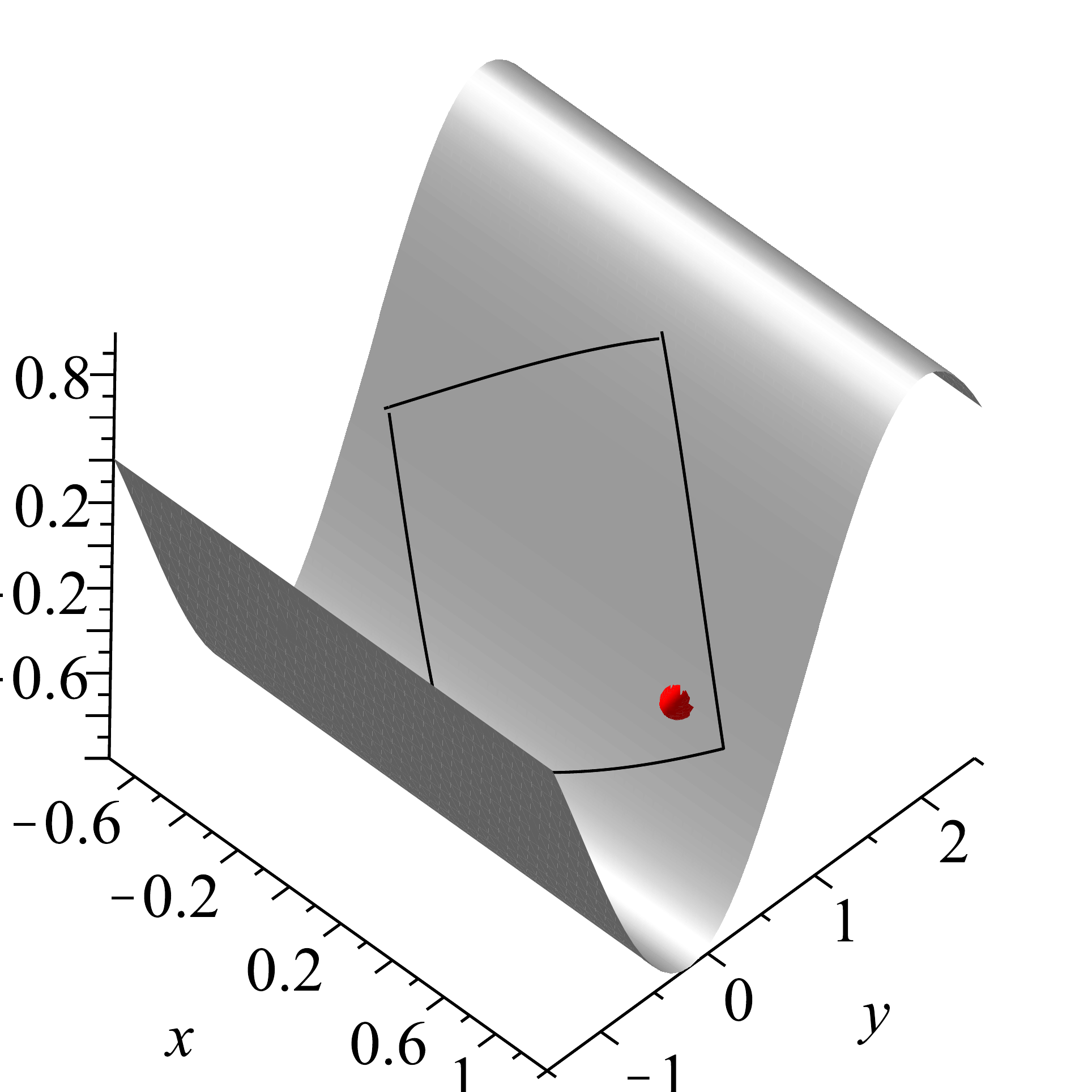}\includegraphics[width=0.5\columnwidth]{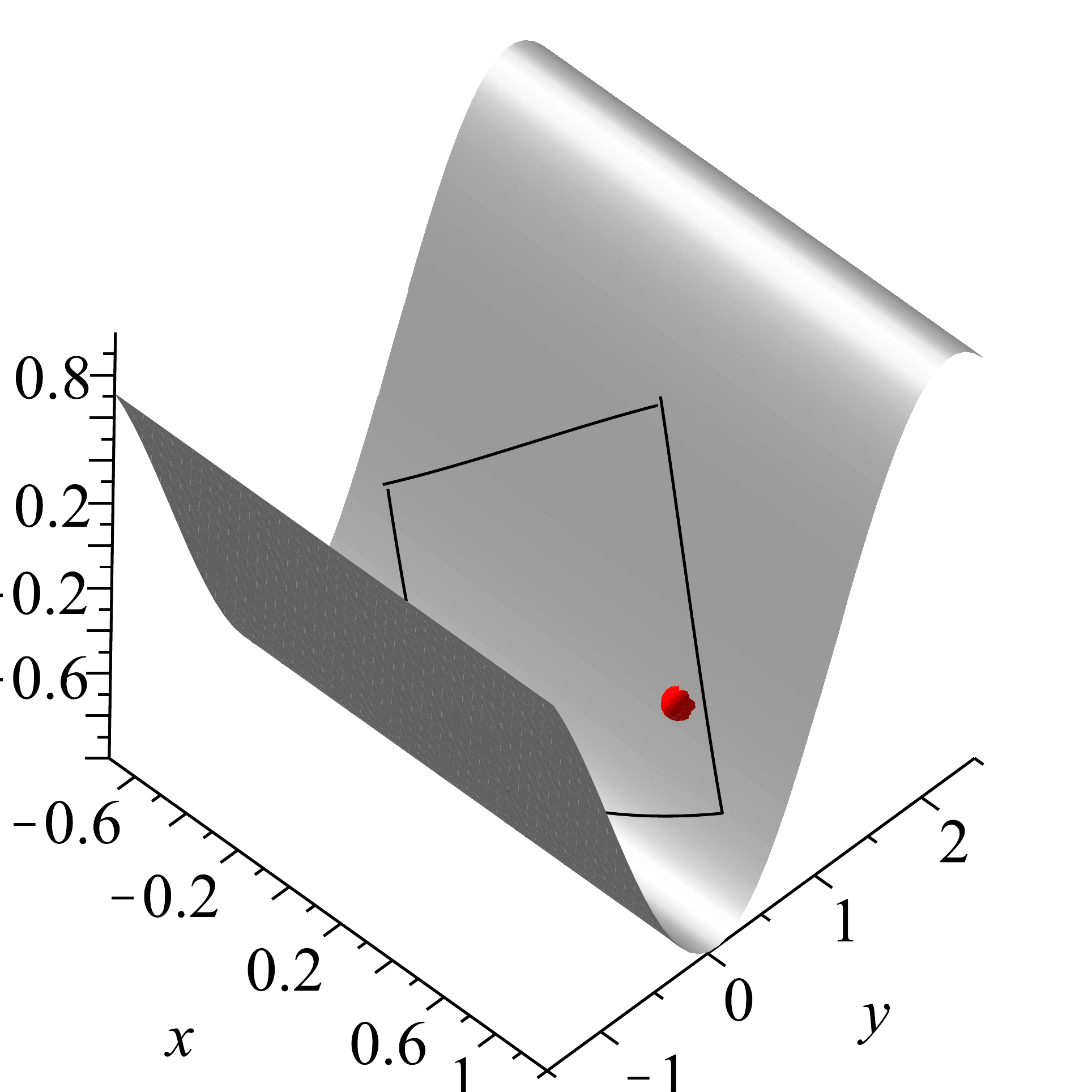}
\par\end{centering}

\begin{minipage}[t]{0.5\columnwidth}%
\begin{center}
(c)
\par\end{center}%
\end{minipage}%
\begin{minipage}[t]{0.5\columnwidth}%
\begin{center}
(d)
\par\end{center}%
\end{minipage}

\centering{}\protect\caption{\textcolor{black}{Illustration of four snapshots of a typical non-planar
time-dependent billiard as explained in Eq. \ref{eq:2-7}. Solid black
lines show the boundary.\label{fig. 1}}}
\end{figure}

\par\end{center}

The metric for the surface in Eq.  \ref{eq:2-7} is
\begin{equation}
\begin{array}{cc}
a_{xx}=1+A^{2}k_{x}^{2} & a_{xy}=A^{2}k_{x}k_{y}\\
\cos^{2}\left(k_{x}x+k_{y}y-\omega t\right)\,\,\,\,\, & \cos^{2}\left(k_{x}x+k_{y}y-\omega t\right)\\
\\
a_{yx}=A^{2}k_{x}k_{y} & a_{yy}=1+A^{2}k_{y}^{2}\\
\cos^{2}\left(k_{x}x+k_{y}y-\omega t\right)\,\,\,\,\, & \cos^{2}\left(k_{x}x+k_{y}y-\omega t\right)
\end{array}.\label{eq:2-9}
\end{equation}
By considering that $k$ and $A$ are both real, this metric is a
Riemannian metric. This is because the determinant of the metric is
positive definite, i.e.,
\[
a_{xx}a_{yy}-\left(a_{xy}\right)^{2}=1+A^{2}k^{2}\cos^{2}\left(k_{x}x+k_{y}y-\omega t\right)>0.
\]
For the Lagrangian and the canonical momentums we get
\begin{eqnarray}
L & = & \frac{1}{2}(a_{xx}(x,y;t)\dot{x}^{2}+2a_{xy}(x,y;t)\dot{y}\dot{x}\nonumber \\
 & + & a_{yy}(x,y;t)\dot{y}^{2}),\label{eq:2-10}\\
\nonumber \\
p_{x} & = & \left(a_{xx}(x,y;t)\dot{x}+a_{xy}(x,y;t)\dot{y}\right),\nonumber \\
p_{y} & = & \left(a_{yy}(x,y;t)\dot{y}+a_{yx}(x,y;t)\dot{x}\right).\label{eq:2-11}
\end{eqnarray}

Therefore, the equations of motion are the following (for more details
see the Appendix)
\begin{eqnarray}
\frac{d}{dt}(\frac{\text{\ensuremath{\partial}}}{\text{\ensuremath{\partial}}\dot{x}}(\frac{1}{2}(a_{xx}(x,y;t)\dot{x}^{2}+2a_{xy}(x,y;t)\dot{y}\dot{x}\nonumber \\
+a_{yy}(x,y;t)\dot{y}^{2})))-\frac{\text{\ensuremath{\partial}}}{\text{\ensuremath{\partial}}x}((\frac{1}{2}(a_{xx}(x,y;t)\dot{x}^{2}\nonumber \\
+2a_{xy}(x,y;t)\dot{y}\dot{x}+a_{yy}(x,y;t)\dot{y}^{2})))=0,\nonumber \\
\nonumber \\
\frac{d}{dt}(\frac{\text{\ensuremath{\partial}}}{\text{\ensuremath{\partial}}\dot{y}}(\frac{1}{2}(a_{xx}(x,y;t)\dot{x}^{2}+2a_{xy}(x,y;t)\dot{y}\dot{x}\nonumber \\
+a_{yy}(x,y;t)\dot{y}^{2})))-\frac{\text{\ensuremath{\partial}}}{\text{\ensuremath{\partial}}y}(\frac{1}{2}(a_{xx}(x,y;t)\dot{x}^{2}\nonumber \\
+2a_{xy}(x,y;t)\dot{y}\dot{x}+a_{yy}(x,y;t)\dot{y}^{2}))=0.\label{eq:2-12}
\end{eqnarray}

Next, we explain our numerical methods. We consider the motion of
a particle constrained to a time-varying surface (Eq. \ref{eq:2-7})
with a square boundary. The square boundary is placed along the x
and y axes. The position of the particle is found by solving equations
of motion numerically. Here, we assume that the boundary is fixed
with respect to the time-dependent surface $\Sigma$. When, the particle
collides with the boundary, the velocity vector would reflect with
respect to the boundary and only the normal component of the velocity
would reverse. New components of the velocity are considered as the
initial conditions for the dynamics after the collision.

In the next section, we explore numerical results for the dynamics
of the particle.

\section{Numerical results }

\subsection{Trajectory of a particle}

\begin{figure}
\includegraphics[width=0.45\columnwidth]{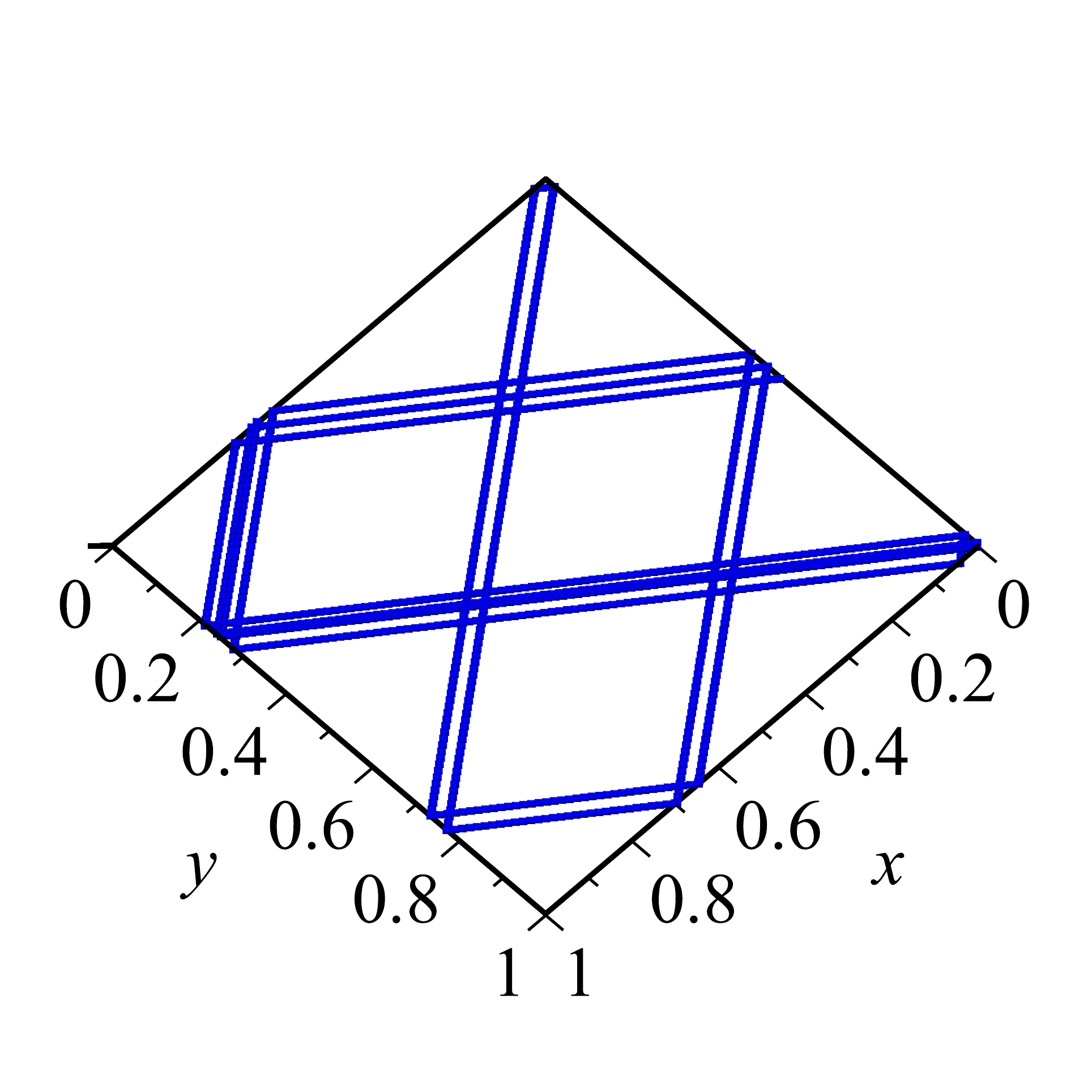}\includegraphics[width=0.45\columnwidth]{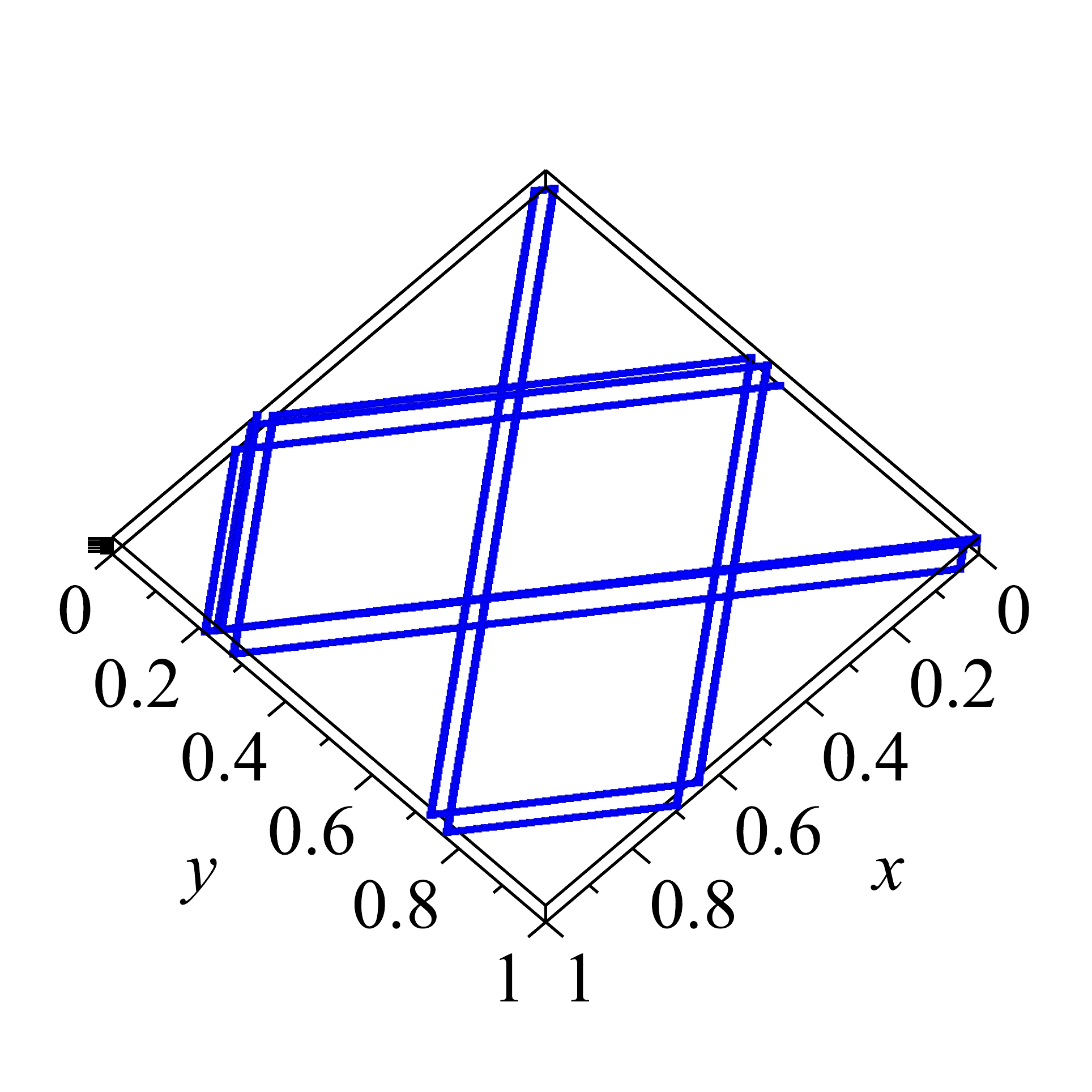}

\begin{minipage}[t]{0.5\columnwidth}%
\begin{center}
(a)
\par\end{center}%
\end{minipage}%
\begin{minipage}[t]{0.5\columnwidth}%
\begin{center}
(b)
\par\end{center}%
\end{minipage}

\includegraphics[width=0.45\columnwidth]{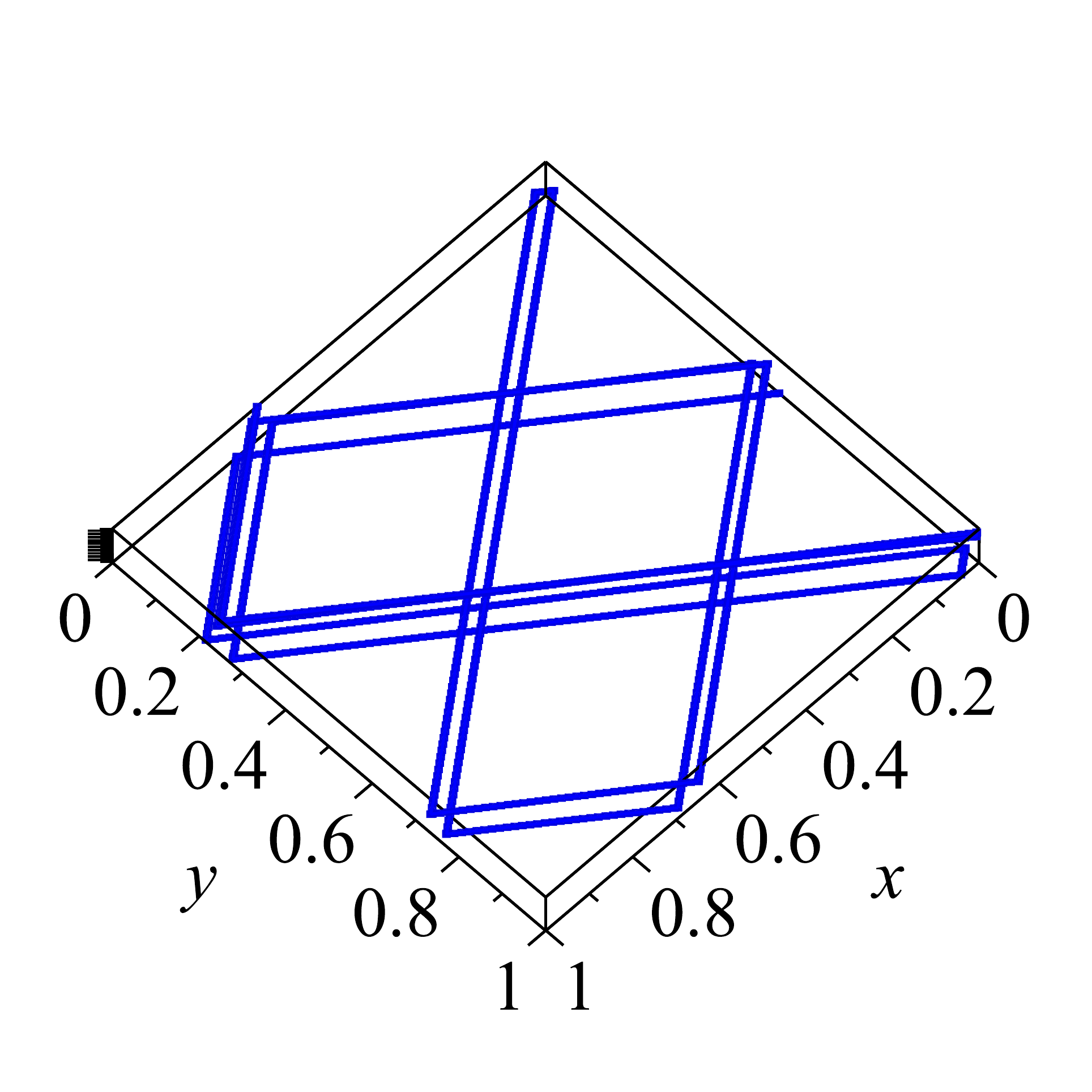}\includegraphics[width=0.45\columnwidth]{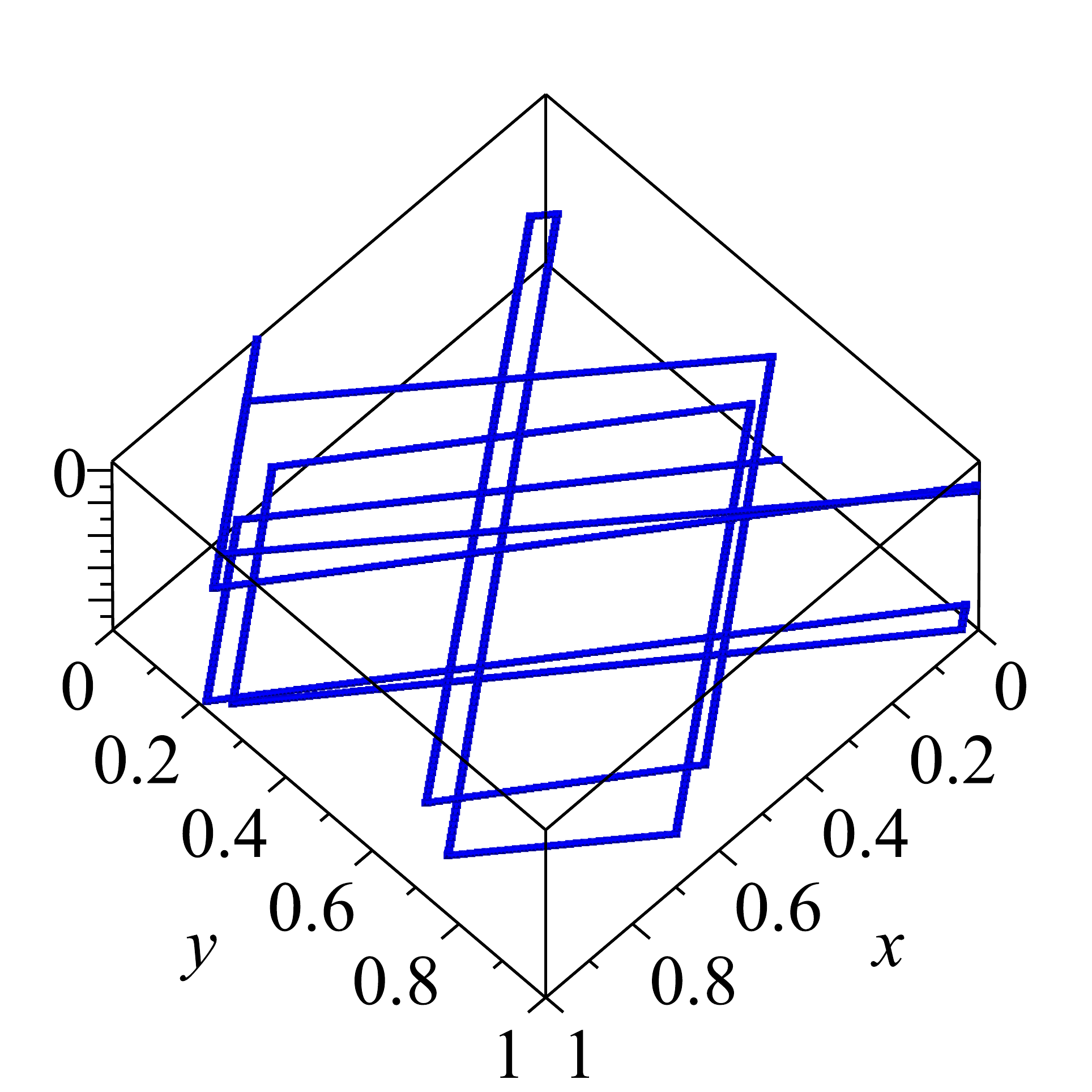}

\begin{minipage}[t]{0.5\columnwidth}%
\begin{center}
(c)
\par\end{center}%
\end{minipage}%
\begin{minipage}[t]{0.5\columnwidth}%
\begin{center}
(d)
\par\end{center}%
\end{minipage}\protect\caption{\textcolor{black}{As $A$ increases, }the trajectory of the particle
becomes more irregular\textcolor{black}{. }We take $A=0.01$, $A=0.05$,
$A=0.1$ and $A=0.5$ respectively\textcolor{black}{.\label{fig. 2}}}
\end{figure}

\begin{figure}
\includegraphics[width=0.45\columnwidth]{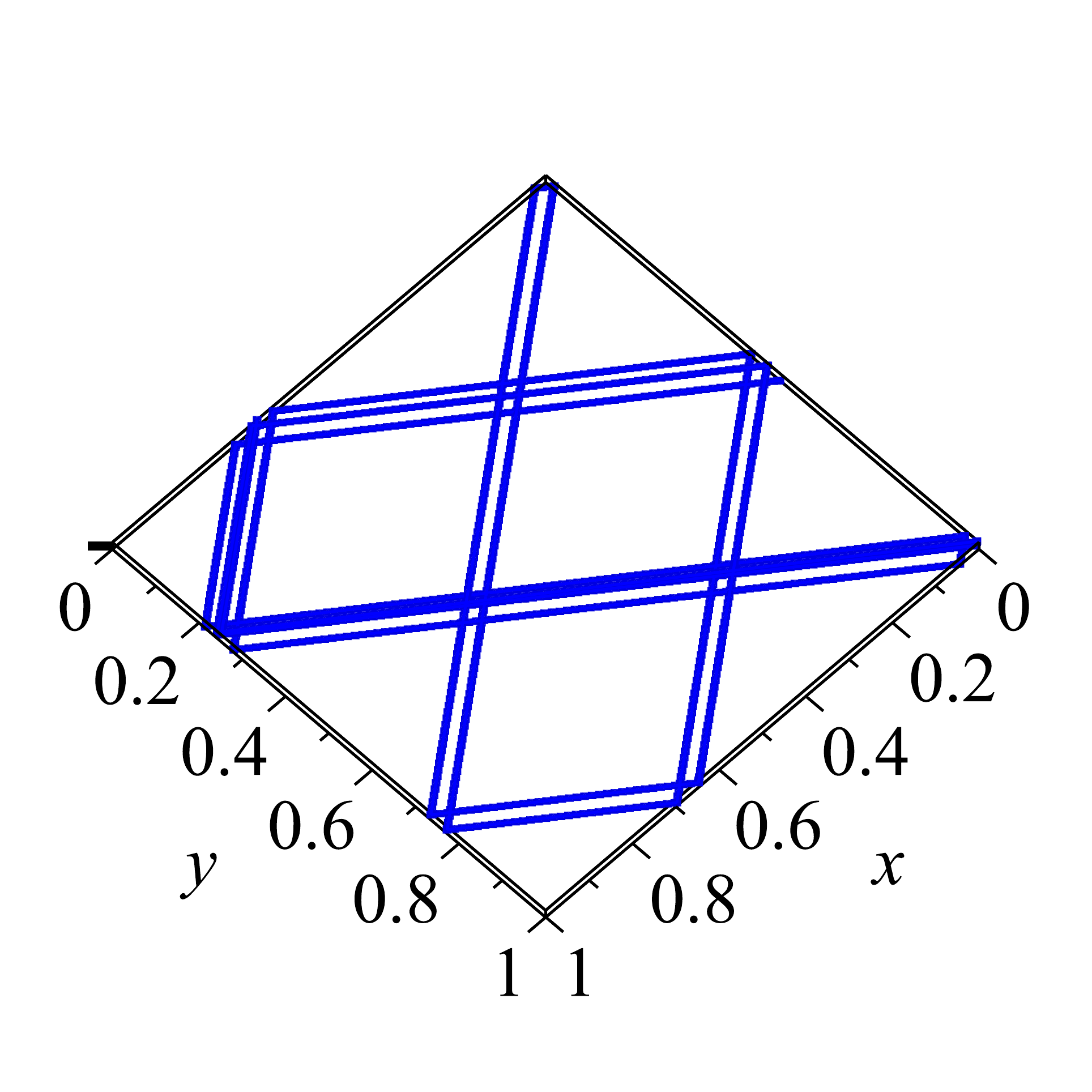}\includegraphics[width=0.45\columnwidth]{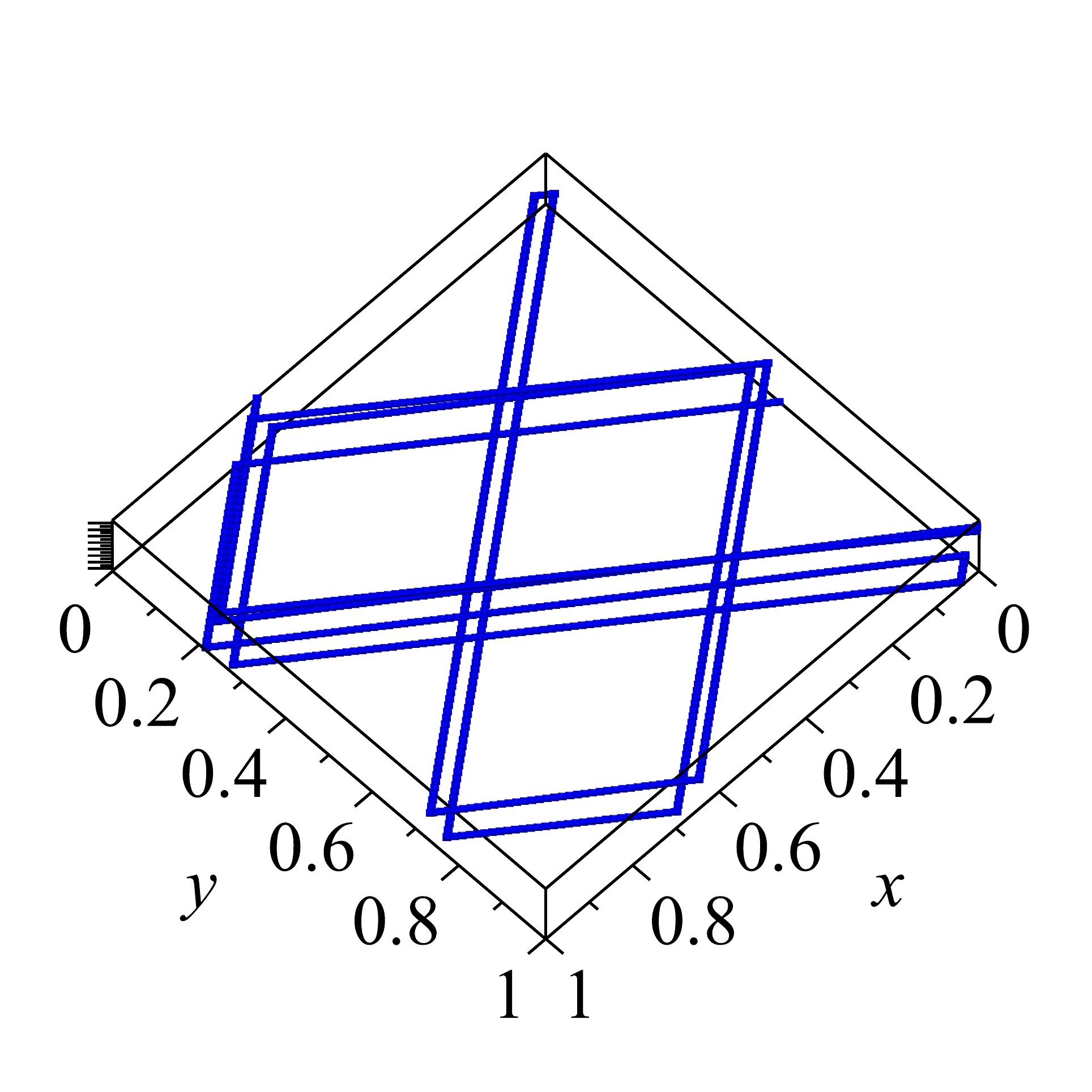}

\begin{minipage}[t]{0.5\columnwidth}%
\begin{center}
(a)
\par\end{center}%
\end{minipage}%
\begin{minipage}[t]{0.5\columnwidth}%
\begin{center}
(b)
\par\end{center}%
\end{minipage}

\includegraphics[width=0.45\columnwidth]{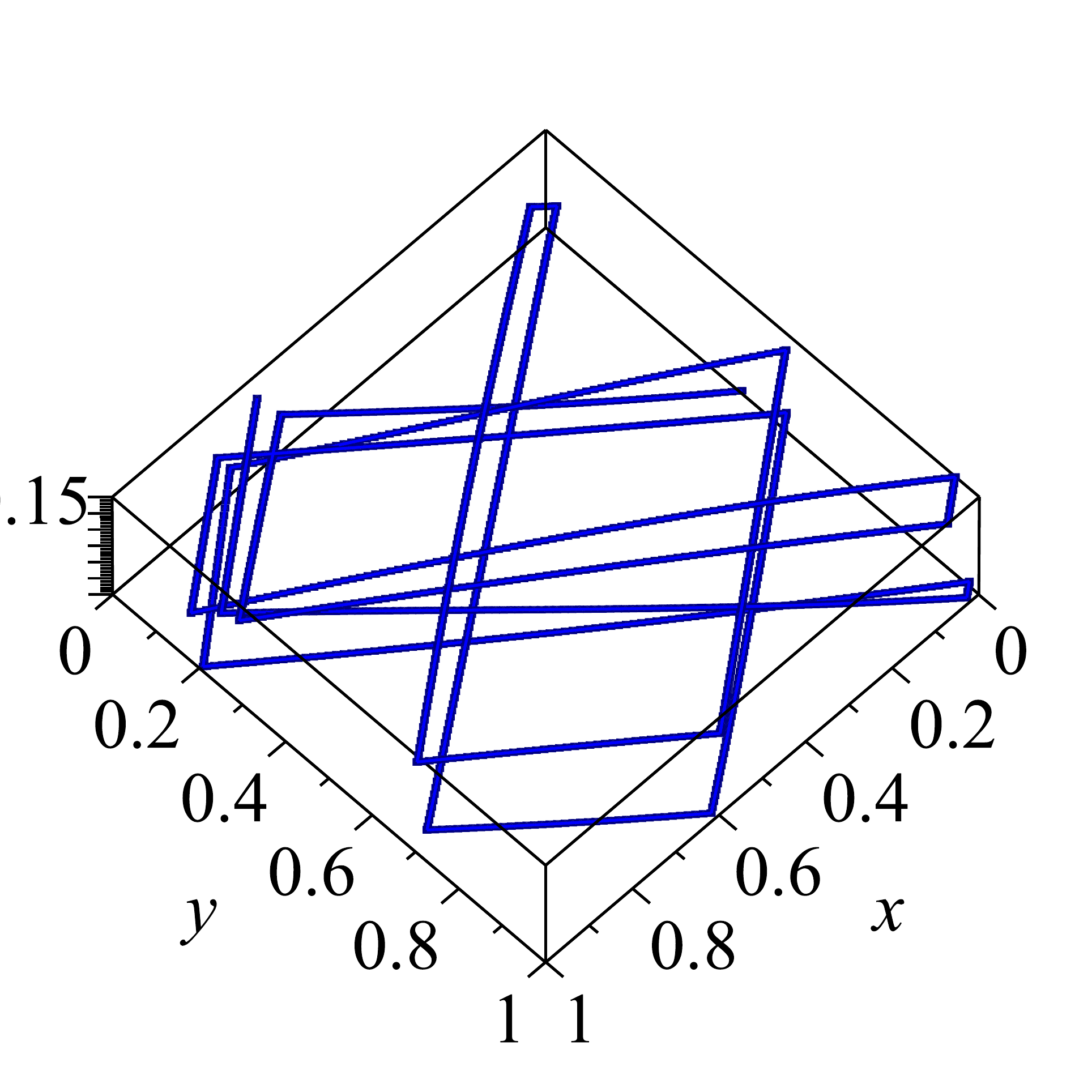}\includegraphics[width=0.45\columnwidth]{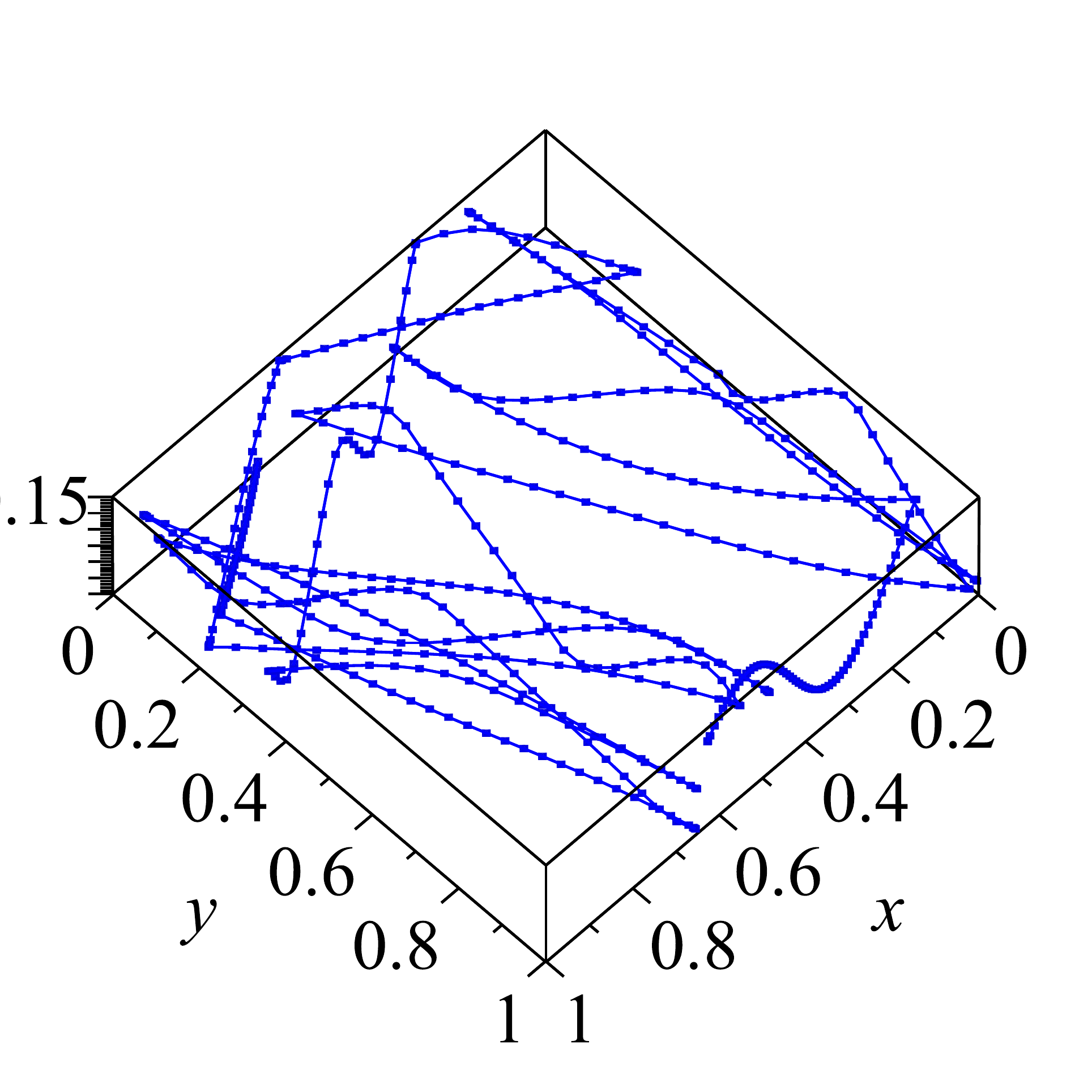}

\begin{minipage}[t]{0.5\columnwidth}%
\begin{center}
(c)
\par\end{center}%
\end{minipage}%
\begin{minipage}[t]{0.5\columnwidth}%
\begin{center}
(d)
\par\end{center}%
\end{minipage}\protect\caption{\textcolor{black}{As $k$ increases, }the trajectory of the particle
becomes more irregular. We take $k=0.01$, $k=0.1$, $k=1$ and $k=10$
respectively.\label{fig. 3}}
\end{figure}

In this subsection, given the Eq. \ref{eq:2-7} as the plane of the
billiard, we investigate behaviors of the trajectory of the particle
in our model. We consider $x_{0}=\frac{2}{3},$ $y_{0}=0,$ $V=1$
and $V_{x}=0.7986355099$ as initial conditions of the particle. The
wave amplitude, $A$, and the wave vector, $k$, control the behavior
of the trajectory. We display the trajectory in the three-dimensional
Euclidean space. Figures \ref{fig. 2} and \ref{fig. 3} illustrate
the trajectories and show how $A$ and $k$ would change the trajectory.

Figure \ref{fig. 2} displays the effect of the wave amplitude as
a control parameter. In Fig. \ref{fig. 2}, we take $A=0.01$, $A=0.05$,
$A=0.1$ and $A=0.5$ respectively. Other parameters; including the
magnitude of the wave vector, the wave velocity, the wave direction
and the time step are
\[
k=0.1,V_{wave}=1,\theta_{wave}=\frac{\pi}{5},\text{\ensuremath{\Delta}}t=\frac{1}{80}.
\]

As the amplitude of the wave increases, the billiard changes from
the familiar flat square billiard to a non-planar square billiard.
When, $A$ is small but non-zero, we see that the trajectory is close
to the periodic orbit of the flat billiard (Fig. \ref{fig. 2} (a)).
As the amplitude of the wave increases, the shape of the trajectory
becomes more irregular. In Fig. \ref{fig. 2} (d) it becomes irregular. 

Figure \ref{fig. 3} displays the effect of increasing the magnitude
of the wave vector. In Fig. \ref{fig. 3}, we take $k=0.01$, $k=0.1$,
$k=1$ and $k=10$ respectively. Other parameters like the wave amplitude,
the wave velocity magnitude, the wave direction and the time step
are 
\[
A=0.15,V_{wave}=1,\theta_{wave}=\frac{\pi}{5},\text{\ensuremath{\Delta}}t=\frac{1}{80}.
\]

The influence of the changes in the wave vector is similar to the
changes in the wave amplitude. As the magnitude of the wave vector
increases, the wave length would decreases. Then, the number of dips
and peaks in the billiard region would increase at each instant of
time. If the wave length is significantly greater than the dimensions
of the billiard then the billiard is almost planar. Thus, as the wave
length decreases, the trajectory of the point particle changes from
the planar periodic orbit.

These diagrams illustrate that each of characteristics of the wave
individually can  influence the shape of trajectories. Therefore,
it is hard to explain the behavior of a trajectory when all of these
parameters have an effective role. 

Furthermore, the direction of wave fronts with respect to the boundary
of the billiard is also an important parameter. A typical example
is the parallel case, when the wave vector is parallel with x or y
axes, then the system has a symmetry plane. This symmetry plane is
placed in the middle of the billiard and is parallel with the wave
vector. But when the wave vector is along another direction, symmetries
of the billiard would change. The net symmetry of the plane and the
boundary of the billiard affects the motion of the particle. Preservation
of symmetries determines the number of constants of motion and the
number of independent constants of motion determines the dimensions
of the subspace of the trajectory in the phase space. If there are
N independent constants of motion, the motion is restricted to a subspace
with dimension 4-N in phase space. (Notice that, when the system is
not conservative, the energy is not a constant of motion. ). We will
explain this in Sec. III. C.

\subsection{Phase space}

\begin{figure}
\centering{}\includegraphics[width=0.55\columnwidth]{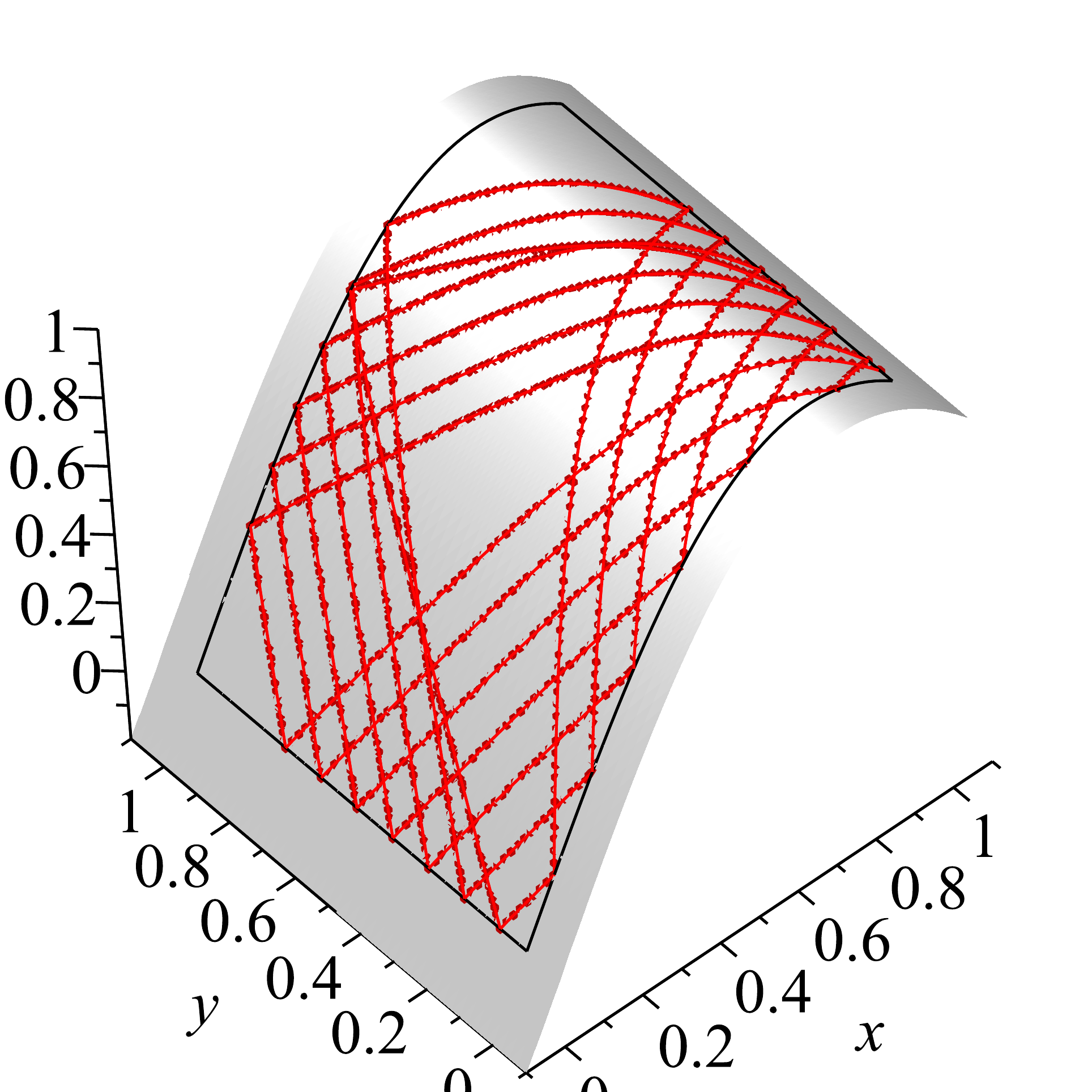}\protect\caption{Illustration of the trajectory of a particle on a typical static billiard.
The initial conditions are $x_{0}=1,y_{0}=\frac{1}{3},V_{x}=-.7193398005,V_{y}=0.6946583703$
and the other parameters used are $A=1,k=2,V_{wave}=0$ and $\theta_{wave}=0$.
\label{fig. 4}}
\end{figure}

In this section, we study the dynamical behavior of the static case
of this billiard. In this case, the wave velocity is zero. The wave
velocity can be obtained by $\frac{\omega}{k}=V_{wave}$, thus, in
the static case $\omega$ vanishes. Using the Eq. \ref{eq:2-7}, the
third component of the surface is obtained as $y^{3}=A\sin\left(k_{x}x+k_{y}y\right)$.
Therefore, the billiard is a static sinusoidal surface with a square
boundary. The Poincare section of the phase space of the static case
is investigated in order to obtain its dynamical behavior. In Fig.
\ref{fig. 4}, the trajectory of a particle in a typical static billiard
is displayed. 

\begin{figure}
\includegraphics[width=0.5\columnwidth]{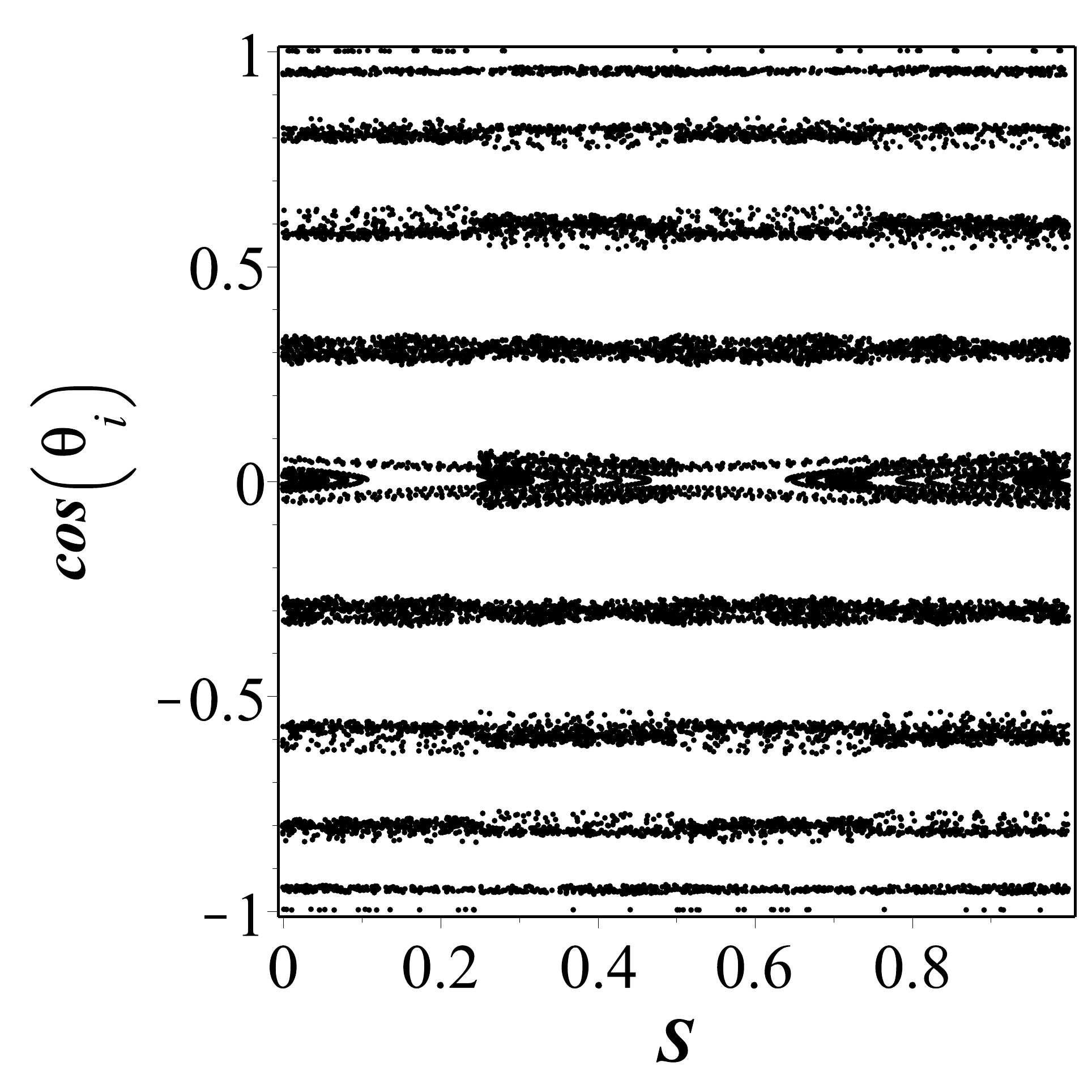}\includegraphics[width=0.5\columnwidth]{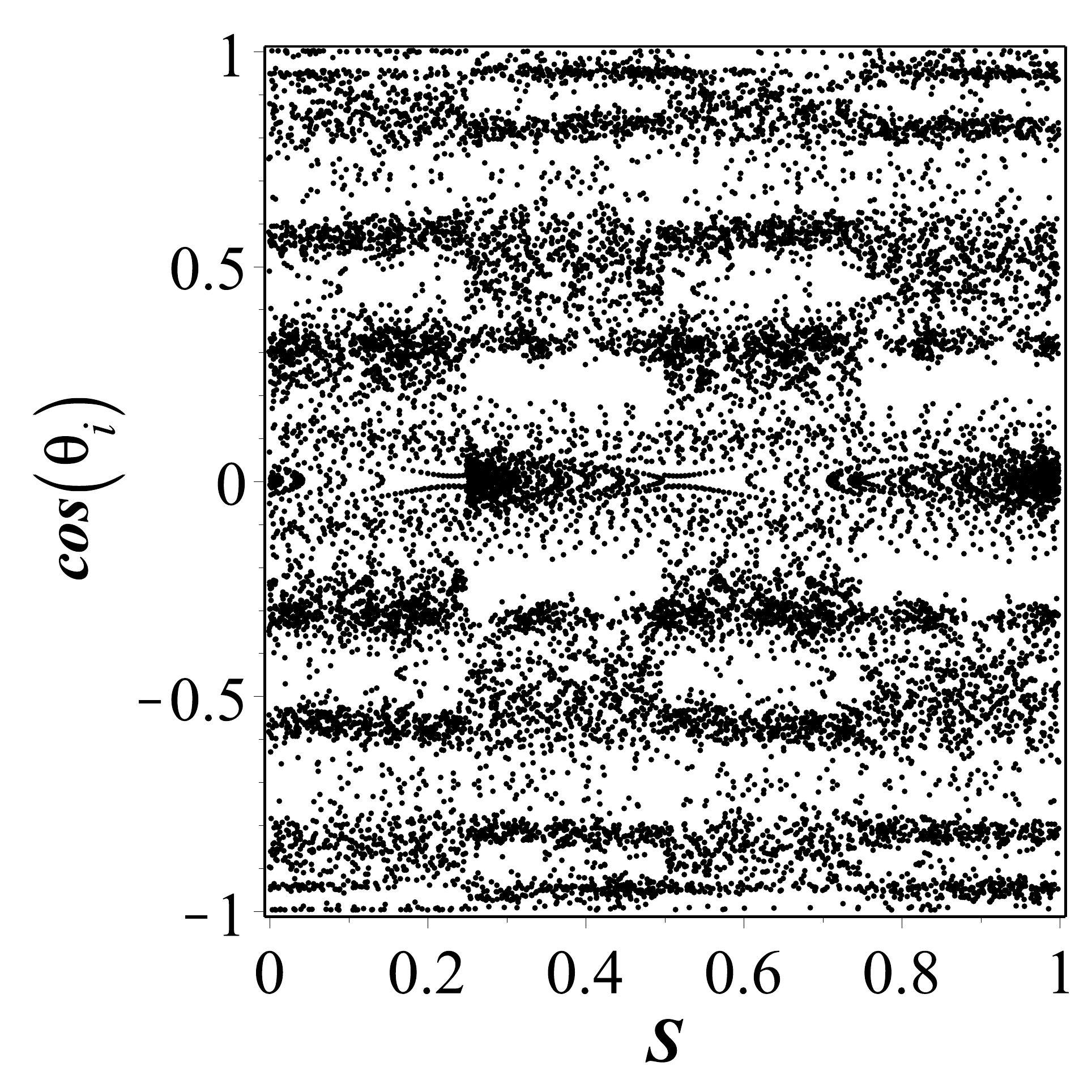}

\begin{minipage}[t]{0.5\columnwidth}%
\begin{center}
(a)
\par\end{center}%
\end{minipage}%
\begin{minipage}[t]{0.5\columnwidth}%
\begin{center}
(b)
\par\end{center}%
\end{minipage}

\begin{centering}
\includegraphics[width=0.5\columnwidth]{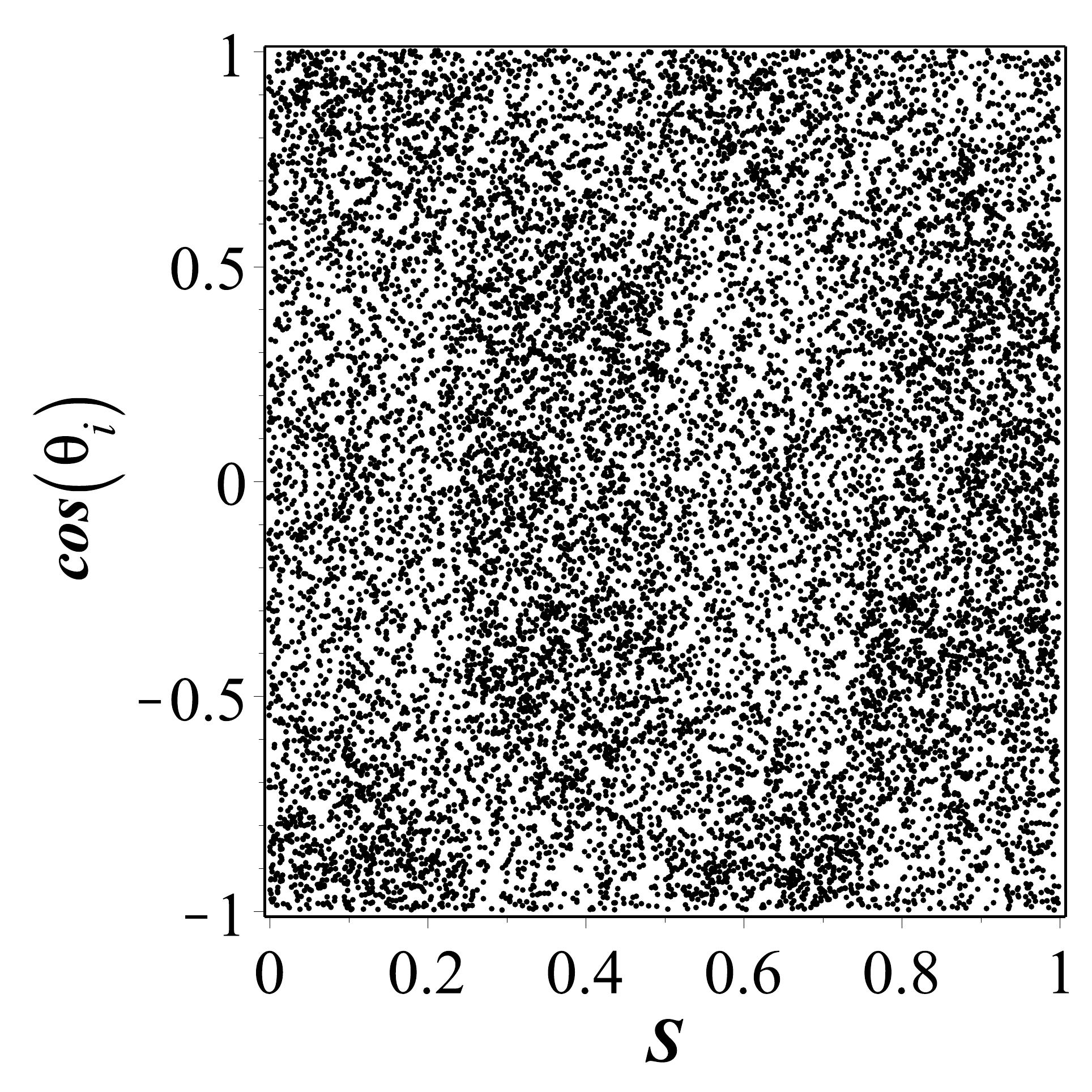}\includegraphics[width=0.5\columnwidth]{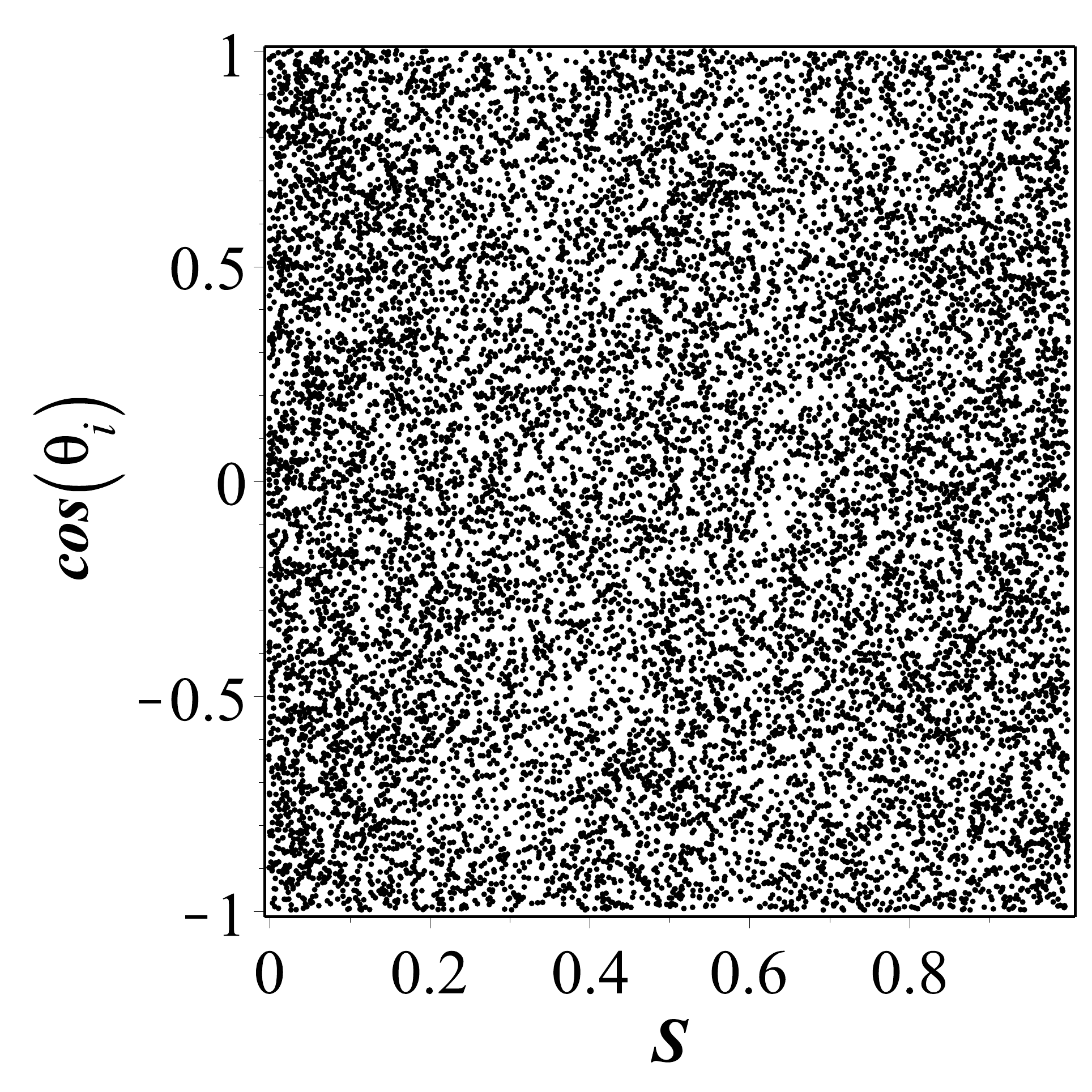}
\par\end{centering}

\centering{}%
\begin{minipage}[t]{0.5\columnwidth}%
\begin{center}
(c)
\par\end{center}%
\end{minipage}%
\begin{minipage}[t]{0.5\columnwidth}%
\begin{center}
(d)
\par\end{center}%
\end{minipage}\protect\caption{The reduced phase space illustrates that, by increasing the amplitude
of the wave, the dynamics of the billiard becomes chaotic.\textcolor{red}{{}
}The wave amplitude and wave vector used are (a) $k=1.5$ and $A=0.05$;
(b) $k=1.5$ and $A=0.1$; (c) $k=1.5$ and $A=0.25$; and (d) $k=1.5$
and $A=1$. \label{fig. 5}}
\end{figure}

\begin{figure}
\includegraphics[width=0.5\columnwidth]{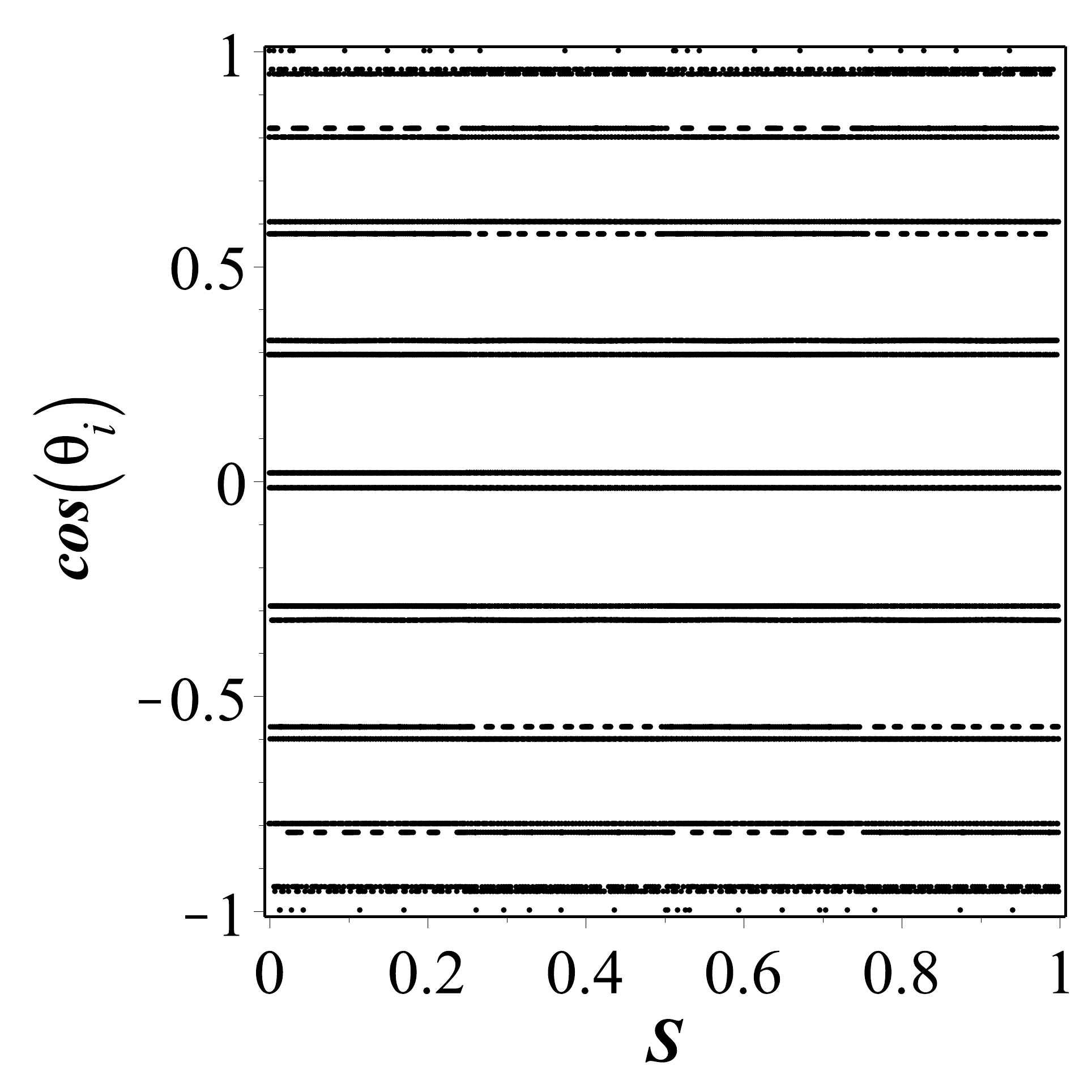}\includegraphics[width=0.5\columnwidth]{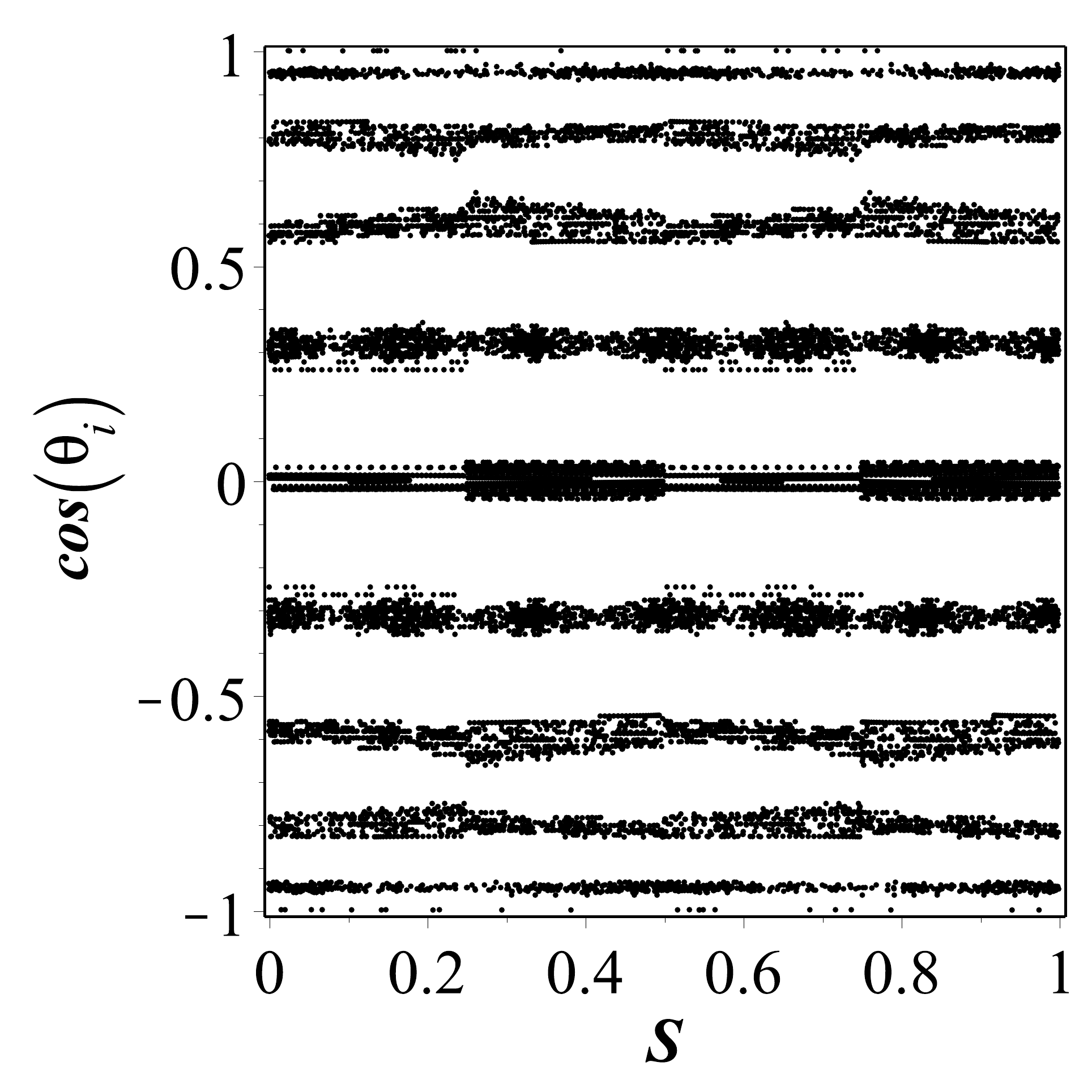}

\begin{minipage}[t]{0.5\columnwidth}%
\begin{center}
(a)
\par\end{center}%
\end{minipage}%
\begin{minipage}[t]{0.5\columnwidth}%
\begin{center}
(b)
\par\end{center}%
\end{minipage}

\begin{centering}
\includegraphics[width=0.5\columnwidth]{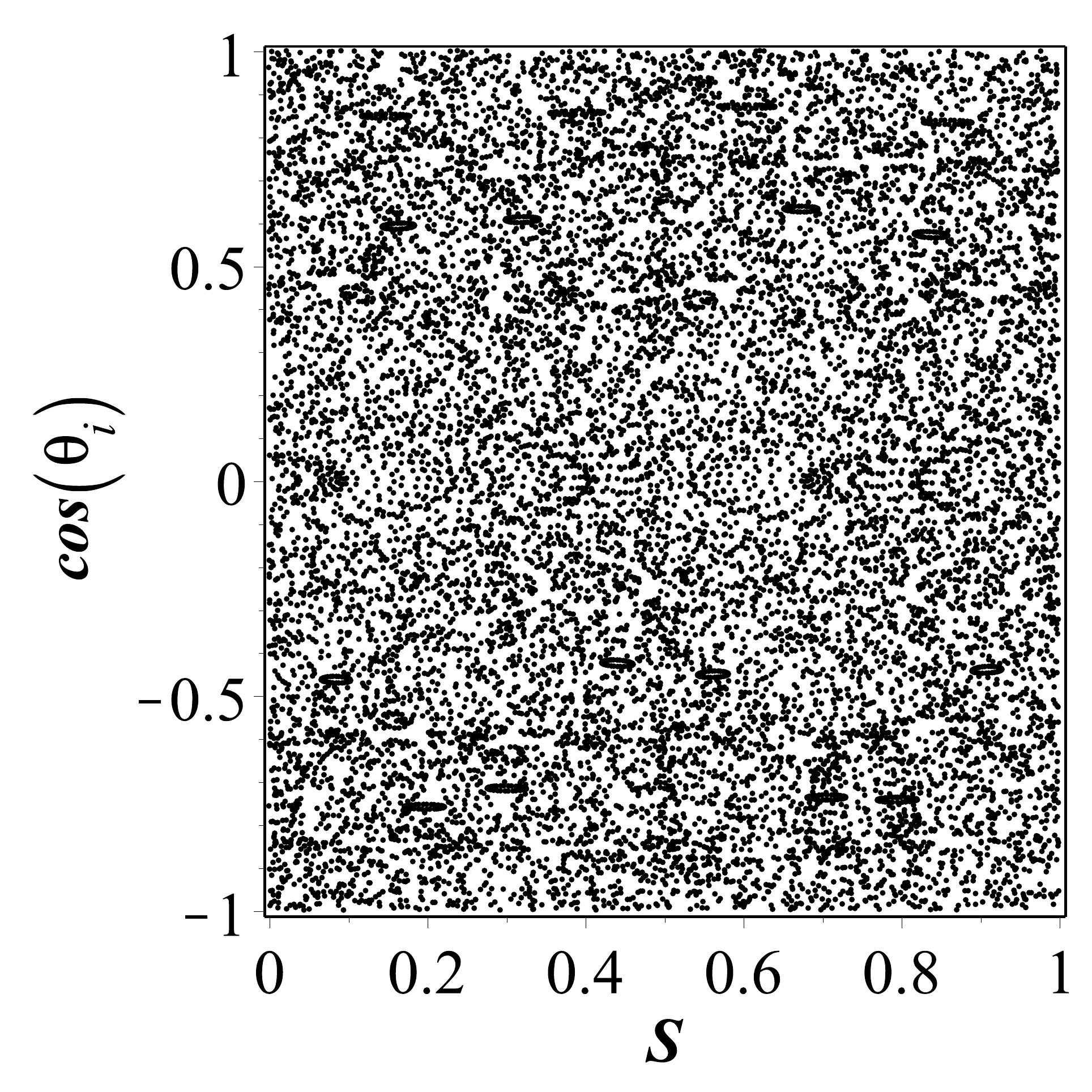}\includegraphics[width=0.5\columnwidth]{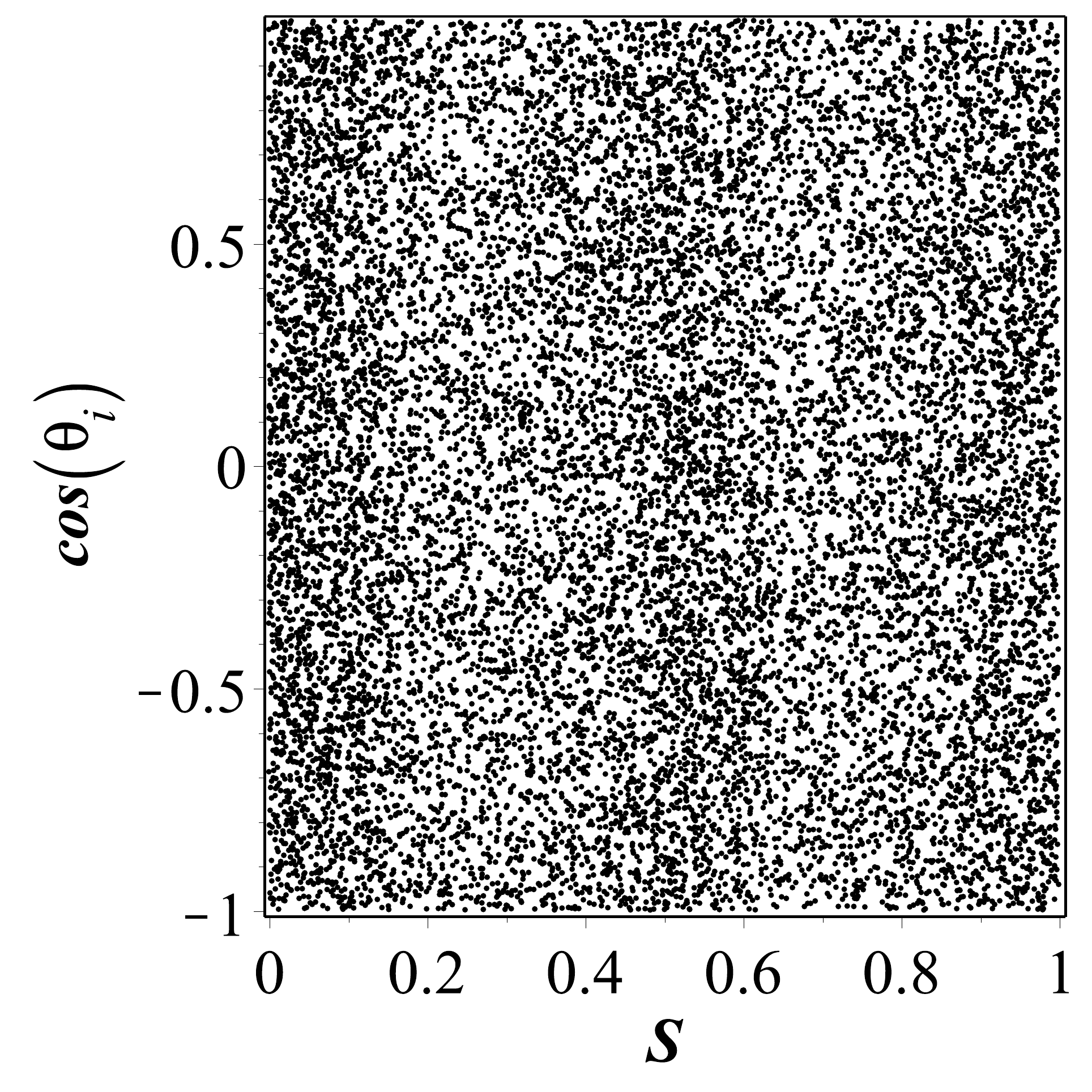}
\par\end{centering}

\centering{}%
\begin{minipage}[t]{0.5\columnwidth}%
\begin{center}
(c)
\par\end{center}%
\end{minipage}%
\begin{minipage}[t]{0.5\columnwidth}%
\begin{center}
(d)
\par\end{center}%
\end{minipage}\protect\caption{The reduced phase space illustrates that, by increasing the magnitude
of the wave vector, the dynamics of the billiard becomes chaotic.
The wave amplitude and wave vector used are (a) $k=0.01$ and $A=1$;
(b) $k=0.1$ and $A=1$; (c) $k=0.5$ and $A=1$; and (d) $k=2$ and
$A=1$.\textcolor{red}{{} }\label{fig. 6}}
\end{figure}

The selection of variables of the reduced phase space is composed
of arc length of each collision point $S\in\left[0,1\right]$ and
the cosine of the angle between the incident velocity vector and the
boundary. In Figs. \ref{fig. 5} and \ref{fig. 6} the reduced phase
spaces are displayed for different static systems when the wave direction
is $\frac{\pi}{5}$. We choose the amplitude of the wave and the magnitude
of the wave vector as control parameters. Each reduced phase space
is built using 30 initial conditions for the first 500 collisions
with the boundary. Figure \ref{fig. 5} shows how the wave amplitude
affects the phase space diagram. 

A particle in non-planar billiard has two degree of freedom and hence,
the phase space is four-dimensional. In the static situation, the
energy is a constant of motion. In the planar billiard, the momentum
is a constant of motion too. While in the non-planar case the symmetry
of the billiard and as a consequence the momentum preservation break
down. Figure \ref{fig. 5} verifies this and shows that, the role
of the growth of the wave amplitude in increasing the chaotic sea.

Next, the effect of increasing the magnitude of the wave vector (or
decreasing the wave length) on the dynamics of the static system is
shown in Fig. \ref{fig. 6}. 

It shows that, as the magnitude of the wave vector increases,  stochastic
points increase. This illustrates the transition of the dynamics of
the billiard from integrable to chaotic.

As a result, by increasing the wave amplitude or the magnitude of
the wave vector, the billiard deviates from the planar case. As the
wave surface emerges in the interior of the square billiard, the symmetries
of the billiard and the number of constants of motion changes. If
the number of constants of motion decreases, then the dynamics of
the billiard won't be integrable anymore. 

In following subsection, we show that in some cases more than one
constant of motion exists.

\subsection{Behavior of the energy growth }

The main goal of this section is to show that two different behaviors
of the particle energy growth are possible in this model. Furthermore,
by considering their corresponding reduced phase plots, we study the
application of the LRA conjecture for this model. 

\begin{figure}
\begin{centering}
\includegraphics[width=0.5\columnwidth]{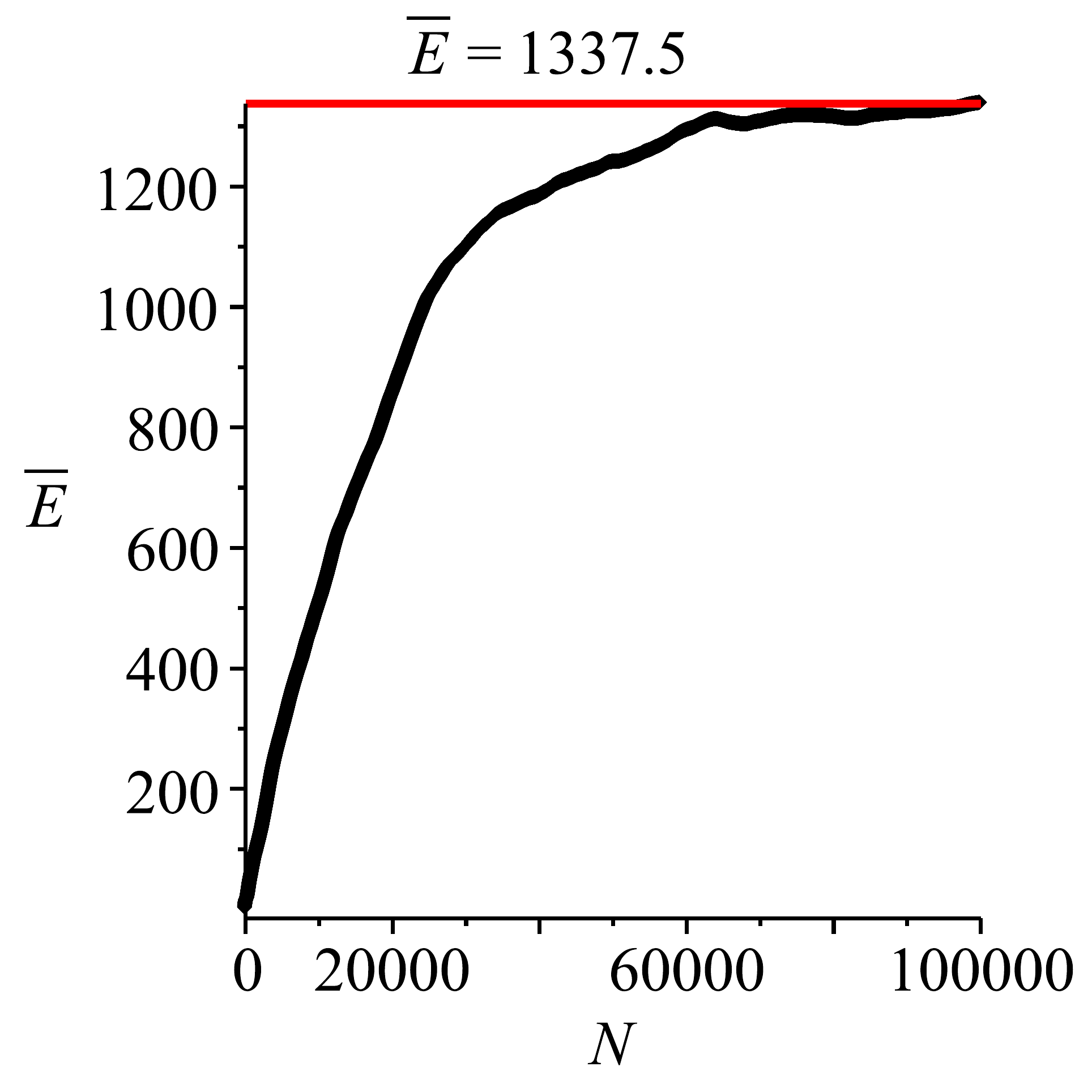}\includegraphics[width=0.45\columnwidth]{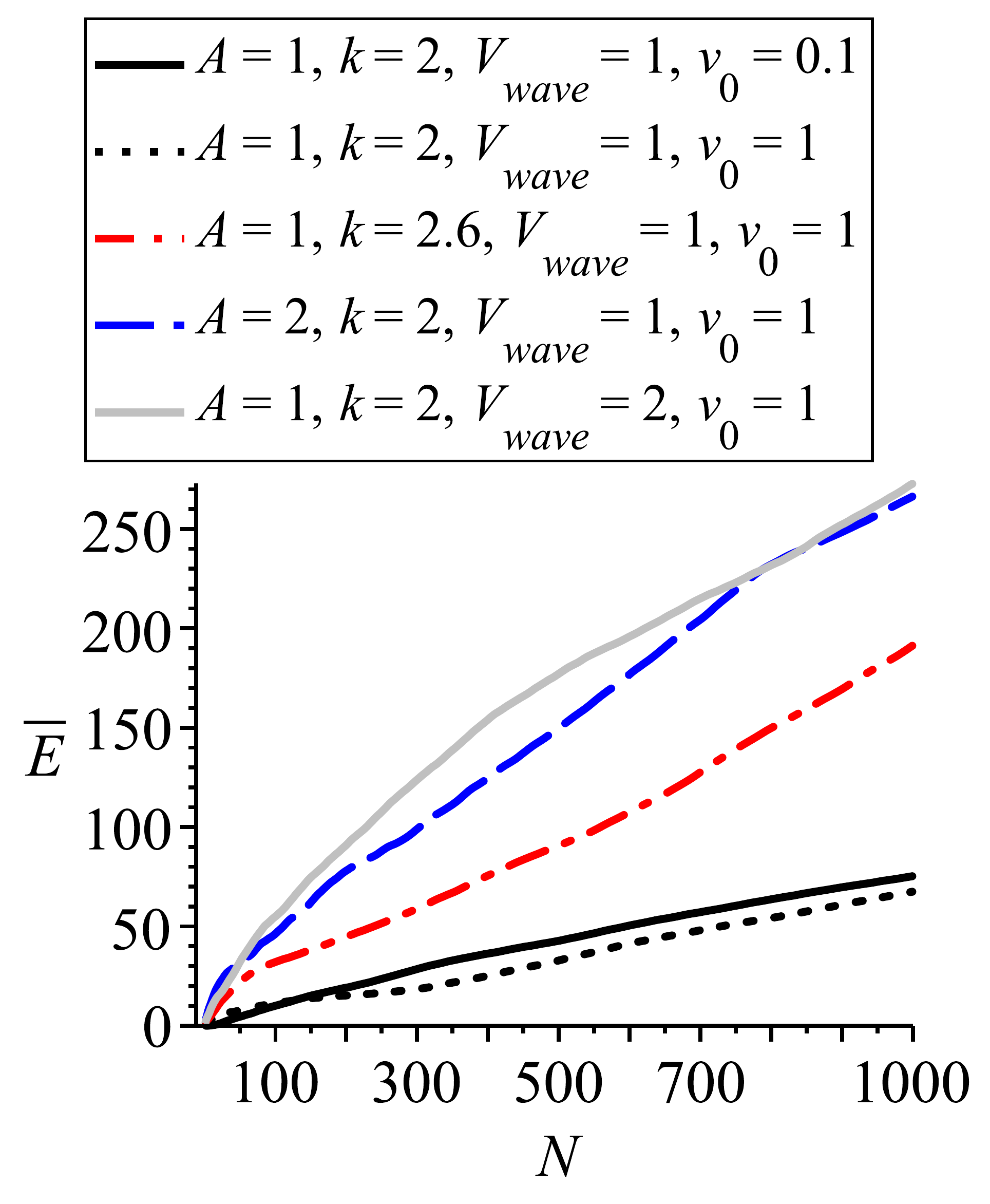}
\par\end{centering}

\begin{minipage}[t]{0.5\columnwidth}%
\begin{center}
(a)
\par\end{center}%
\end{minipage}%
\begin{minipage}[t]{0.5\columnwidth}%
\begin{center}
(b)
\par\end{center}%
\end{minipage}

\protect\caption{For $\theta_{wave}=\frac{\pi}{5}$, the average energy as a function
of the number of collisions is calculated for an ensemble of 30 particles.
(a) When the dynamics of the static case is chaotic, the average energy
grows and then saturates. (b) For comparison, the behavior of energy
growth for different values of the parameters of the system is simulated.\textcolor{cyan}{{}
}\label{fig. 7}}
\end{figure}

The LRA conjecture states that ``chaotic dynamics of a billiard with
a fixed boundary is a sufficient condition for the Fermi acceleration
in the system when a boundary perturbation is introduced''. The perturbation
as a time dependent change in the shape of a boundary has been investigated
in \cite{gelfreich2011robust,Carvalho2006,Leonel2004,Leonel2010,Livorati2008,Oliveira2010,Oliveira2010a,oliveira2012scaling,batistic2014exponential}.
Here, we extend these studies to a billiard with time-dependent plane.
Thus, in this subsection the behavior of the average energy of the
particle as a function of the number of collisions is studied for
a time-dependent non-planar square billiard. Numerical investigations
of the average energy behavior have been performed in two cases, when
the energy grows\textcolor{red}{{} }and then converges to a constant
value, and for a case that the energy remains approximately constant.
The two cases are shown respectively in Figs. \ref{fig. 7} and \ref{fig. 10}.
Results are made in an ensemble of 30 particles for different initial
conditions and characteristics of the wave. Here, we discuss our results
in two separate cases.

\textit{Case 1.} In Fig. \ref{fig. 7}, we consider a situation where
the system exhibits limited energy growth and then it saturates.

Figure \ref{fig. 7} (a) shows that, the average energy increases
for the small number of collisions, and then starts to saturate. The
corresponding phase plot for this case is displayed in Fig. \ref{fig. 6}
(d). The phase plot shows that for this selection of the parameters,
the system has a chaotic dynamics. Therefore, for the small number
of collisions, the results in Fig. \ref{fig. 7} are in agreement
with the LRA conjecture. However, for sufficiently large number of
iterations, our results don't agree with the LRA conjecture and the
average energy converges to the $\bar{E}=1337.5$.

We also compare the results for different values of parameters in
Fig. \ref{fig. 7} (b). An increase in either of the $k$, $V_{wave}$
or $A$ would result in increase of the particle energy.\textcolor{red}{{} }

\begin{figure}
\begin{centering}
\includegraphics[width=0.58\columnwidth]{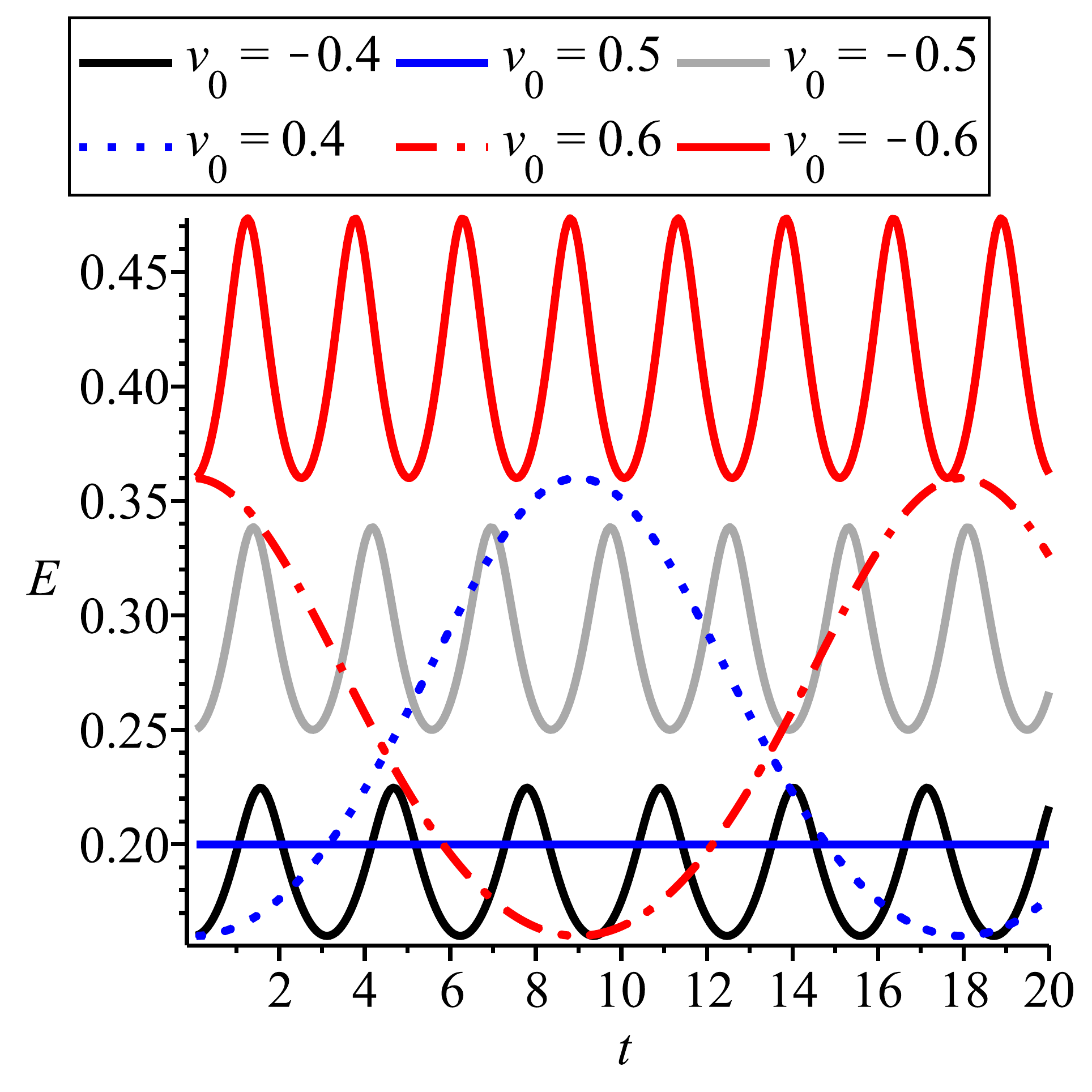}
\par\end{centering}

\protect\caption{The time evolution of the free particle energy for different values
of particle velocity when $\theta_{paricle}=\frac{\pi}{2},$ $\theta_{wave}=\frac{\pi}{2},$
$k=1,$ $A=1$ and $V_{wave}=0.5$. \label{fig. 8}}
\end{figure}

In order to explain the behavior of the energy growth, we consider
the particle on the wave without boundaries. The behavior of the energy
of the free particle as a function of time (Fig. \ref{fig. 8}) explains
why the energy grows and then saturates in Fig. \ref{fig. 7} (a).

The behavior of the particle is better understood in a coordinate
with diagonal metric. When the particle is in the billiard, we don't
use this coordinate system. By considering, $q^{1}=\frac{1}{k}\left(k_{x}x+k_{y}y\right)$,
$q^{2}=\frac{1}{k}\left(-k_{y}x+k_{x}y\right)$, the metric would
be diagonalized. By transforming the coordinates of the particle $\left(x,y\right)$
to the diagonal coordinates $\left(q^{1},q^{2}\right)$, we get
\[
\begin{array}{c}
\begin{array}{cc}
a_{11}=1+A^{2}k^{2}\left(\cos\left(k\,q^{1}-\omega t\right)\right)^{2}\,\,\,\, & a_{12}=0\\
a_{21}=0\,\,\,\, & a_{22}=1
\end{array},\\
\\
L=\frac{1}{2}\left(\left(1+A^{2}k^{2}\left(\cos\left(k\,q^{1}-\omega t\right)\right)^{2}\right)\left(\dot{q}^{1}\right)^{2}+\left(\dot{q}^{2}\right)^{2}\right),\\
\\
\frac{d}{dt}\frac{\partial L}{\partial\dot{q}^{1}}-\frac{\partial L}{\partial q^{1}}=0,\,\,\,\,\,\ddot{q}^{2}=0.
\end{array}
\]
The transformation equations show that, the first component of the
diagonal coordinates is parallel, $v_{q^{1}}=\dot{q}^{1}$, and the
other one is perpendicular to the wave vector, $v_{q^{2}}=\dot{q}^{2}$. 

The equations of motion show that, the perpendicular component of
the free particle velocity remains constant during the motion; while,
the parallel component changes. Therefore, the behavior of the parallel
component of the velocity is responsible for the energy evolution
of the particle.

So, depending on the relative velocity of the particle and the wave,
the energy may increase or decrease with respect to its initial value.

In Fig. \ref{fig. 8}, for all the plots, the initial velocity of
the particle is in the direction of the wave vector. Only in one case
the energy oscillates between the initial value and a lower value.
While, in the other cases, the energy oscillates between its initial
value and a higher value or remains constant.

Consequently, in most of these cases the energy grows. In non-integrable
conditions, it is more probable that the particle collides with the
boundary before its energy reaches its value at the previous reflection.
In other words, it is more probable that the time interval between
two successive collisions isn't multiple of the period of energy evolution.
Since, in most of the energy evolution, energy oscillates between
its initial value and a higher value, it is more probable that the
particle has a higher energy when it reaches the boundary. This is
the reason of energy growth for the small number of collisions in
Fig. \ref{fig. 7}. When, the particle velocity grows enough, the
parallel component, $v_{q^{1}}$ is almost greater than the $V_{wave}$.
There are two possible direction for $v_{q^{1}}$ and each of them
would result in a different energy evolution. One of them oscillates
between its initial value and a higher value of the energy and the
other one oscillates between its initial value of the energy and a
lower value. The balance of these two cases would result in the saturation
behavior of the energy. These cases correspond to $v_{0}=0.6$ and
$v_{0}=-0.6$ in Fig. \ref{fig. 8}. 

\begin{figure}
\begin{centering}
\includegraphics[width=0.58\columnwidth]{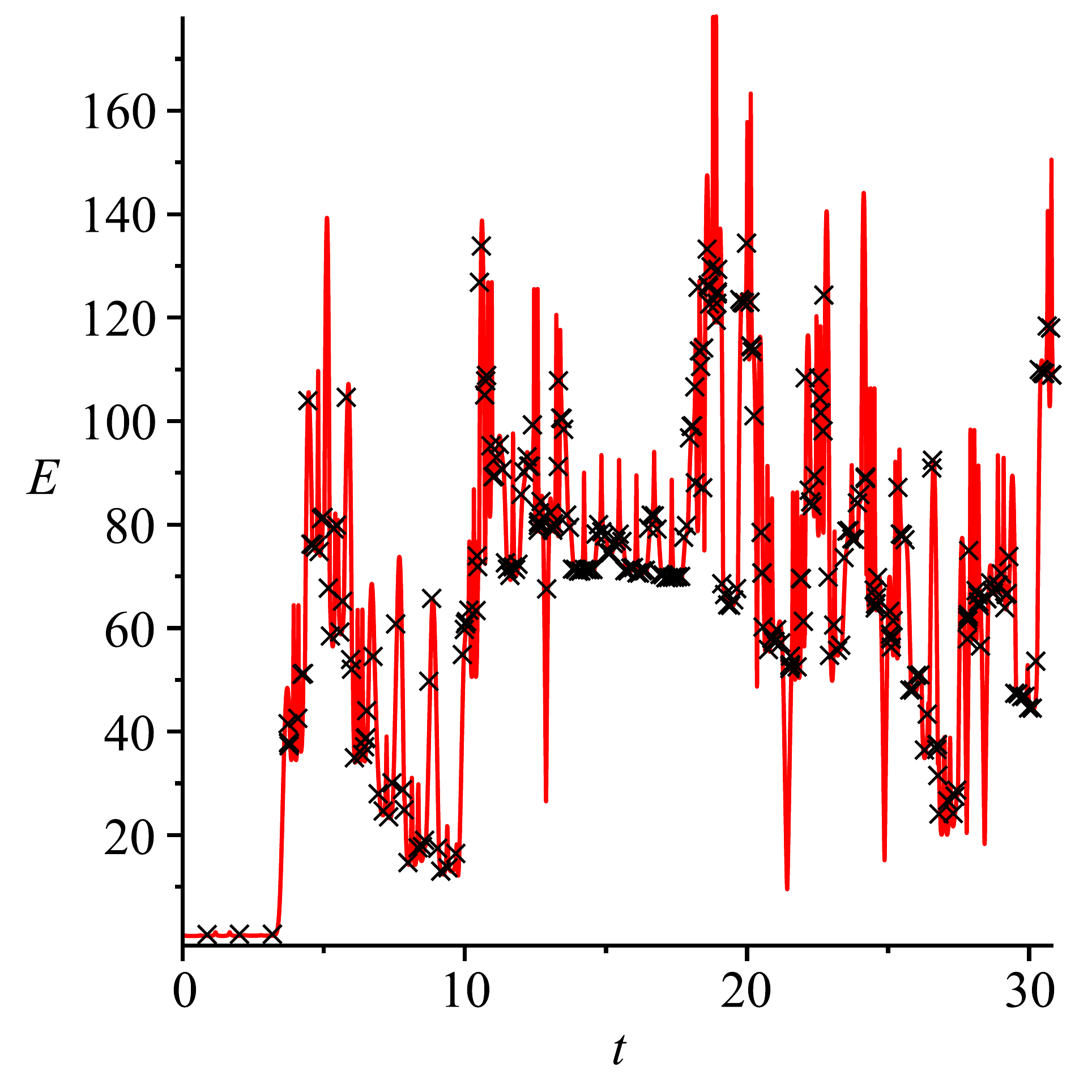}
\par\end{centering}

\protect\caption{The time evolution of a particle energy with initial conditions $x_{0}=1,$
$y_{0}=\frac{1}{3},$ $V_{x}=-.7193398005$ and $V_{y}=0.6946583703$
in a system with $V_{wave}=1,$ $k=5,$ $A=1$ and $\theta_{wave}=\frac{\pi}{5}$.
\label{fig. 9}}
\end{figure}

\begin{figure}
\begin{centering}
\includegraphics[width=0.5\columnwidth]{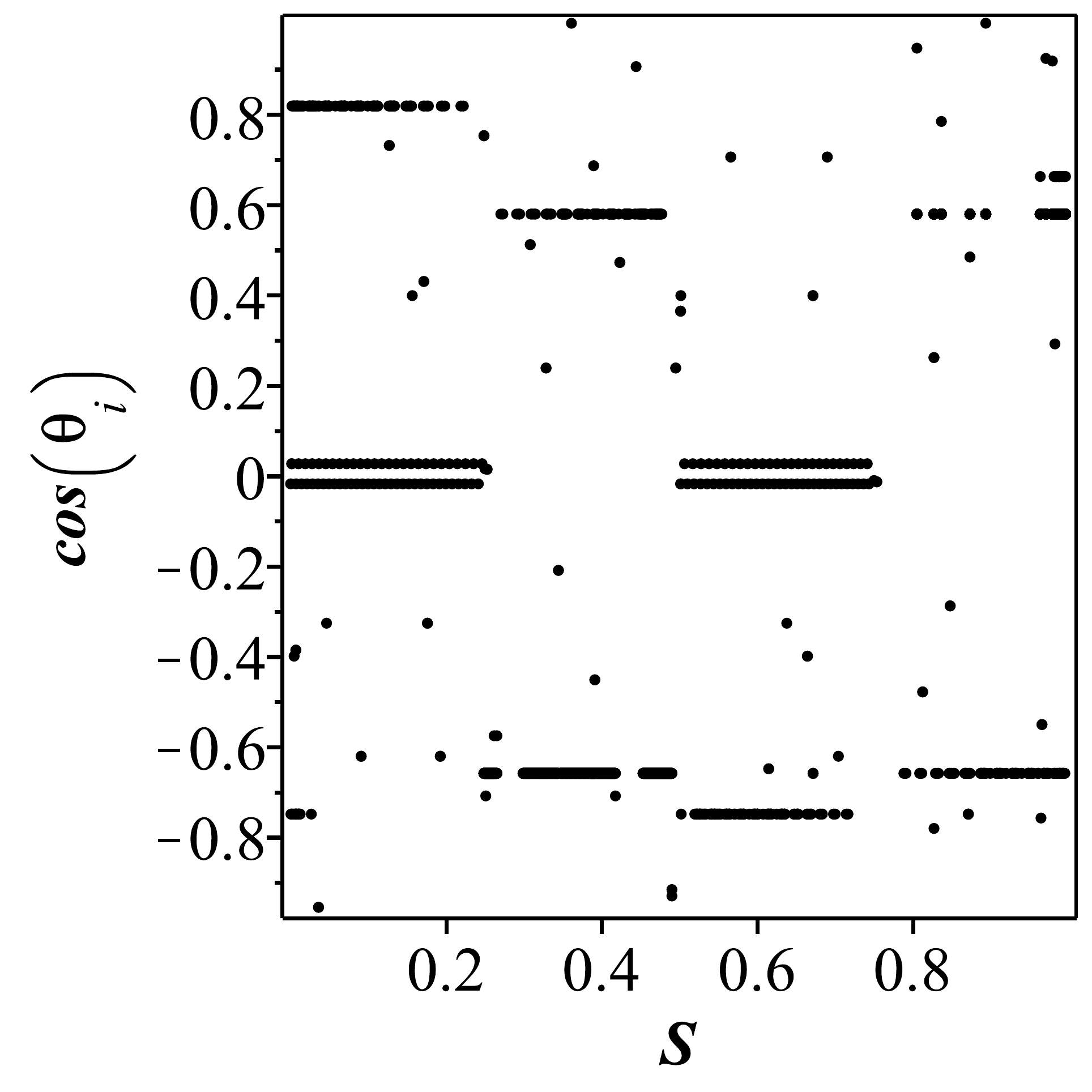}\includegraphics[width=0.5\columnwidth]{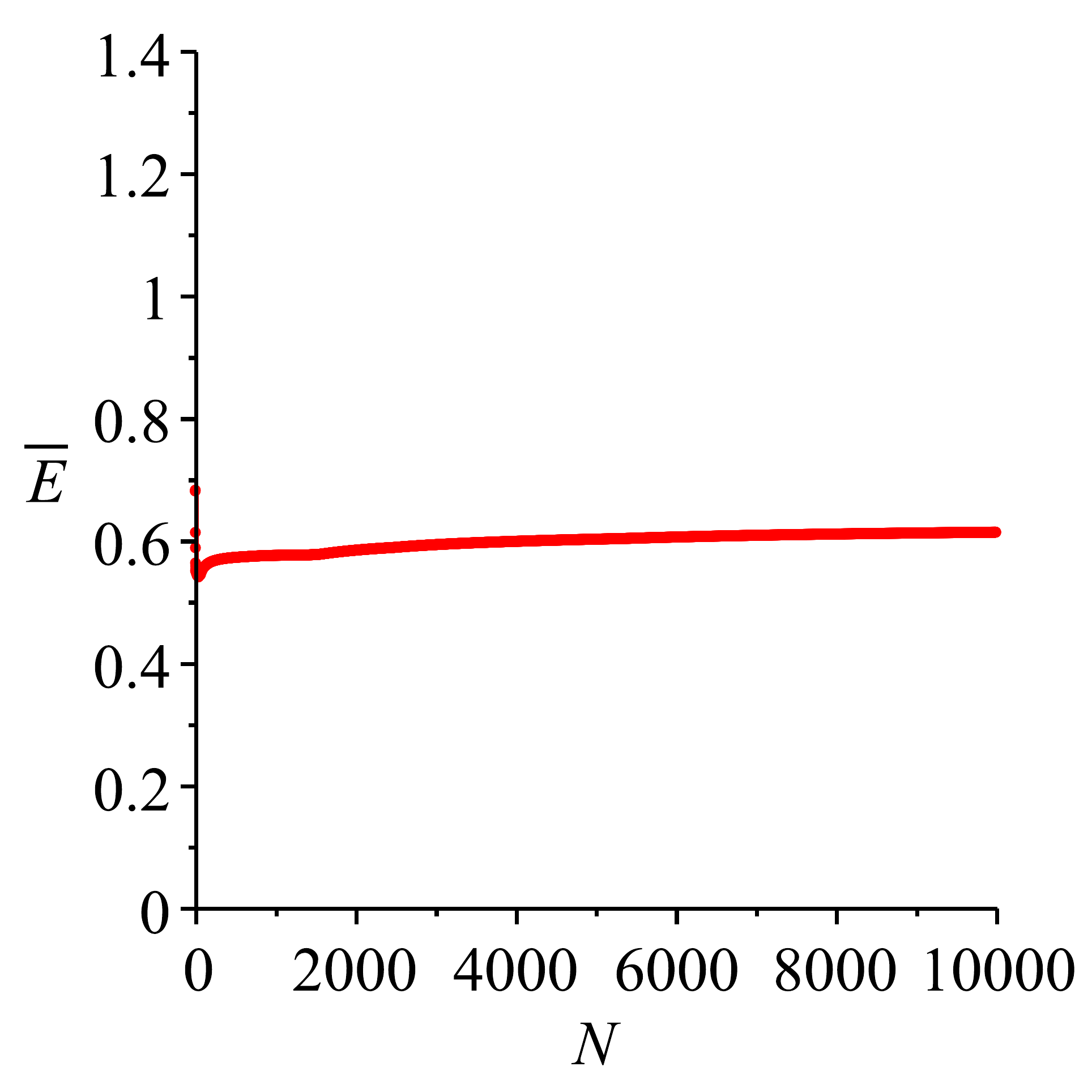}
\par\end{centering}

\begin{centering}
\begin{minipage}[t]{0.5\columnwidth}%
\begin{center}
(a)
\par\end{center}%
\end{minipage}%
\begin{minipage}[t]{0.5\columnwidth}%
\begin{center}
(b)
\par\end{center}%
\end{minipage}
\par\end{centering}

\protect\caption{(a) The reduced phase plot of the static system of $k=1,$ $A=1$
and $\theta_{wave}=0$. (b) The behavior of the average energy for
an ensemble of 30 particles as a function of the number of collisions
in the corresponding time-dependent system. The average energy stays
constant when the dynamics of the static case is integrable. \label{fig. 10}}
\end{figure}

In Fig. \ref{fig. 9}, we present the time evolution of the energy
of a particle for one initial condition during successive impacts
with the boundary in our model. The black crosses specify when the
particle is reflected. As it is seen, the energy of the particle grows\textcolor{red}{{}
}at the beginning of the motion. 

\textit{Case 2.} We now discuss the situation that, the particle average
energy stays near a constant value.

As shown in Fig. \ref{fig. 10} (b), in contrast with the previous
case, the average energy doesn't grow. This figure shows that, even
for the small iteration numbers, the average energy stays near a constant
value which is approximately 0.68. The Fig. \ref{fig. 10} (a) shows
the phase plot of the corresponding static system.

In Fig. \ref{fig. 10}, the wave vector is parallel with two walls
of the boundary. So, the system preserves a kind of symmetry. When
the $\theta_{wave}=0$ , we find that the Eq. \ref{eq:2-7} has the
following form
\begin{eqnarray*}
y^{1} & = & u^{1}=x,\\
y^{2} & = & u^{2}=y,\\
y^{3} & = & A\sin\left(k_{x}x-\omega t\right).
\end{eqnarray*}
Thus, using the Eq. \ref{eq:2-9} the metric is diagonal 
\[
\begin{array}{cc}
a_{xx}=1+A^{2}k_{x}^{2}\cos^{2}\left(k_{x}x-\omega t\right) & a_{xy}=0\\
a_{yx}=0 & a_{yy}=1
\end{array}.
\]
Then, momentums and the Lagrangian are obtained as
\begin{eqnarray*}
p_{x} & = & \left(1+A^{2}k_{x}^{2}\cos^{2}\left(k_{x}x-\omega t\right)\right)\dot{x},\\
p_{y} & = & \dot{y},
\end{eqnarray*}
\[
L=\frac{1}{2}\left(\left(1+A^{2}k_{x}^{2}\cos^{2}\left(k_{x}x-\omega t\right)\right)\dot{x}^{2}+\dot{y}^{2}\right).
\]
using the Lagrangian, equations of motion are written as
\begin{eqnarray*}
\frac{d}{dt}(\frac{\text{\ensuremath{\partial}}}{\text{\ensuremath{\partial}}\dot{x}}(\frac{1}{2}(a_{xx}(x,y;t)\dot{x}^{2}+\dot{y}^{2})))-\\
\frac{\text{\ensuremath{\partial}}}{\text{\ensuremath{\partial}}x}((\frac{1}{2}(a_{xx}(x,y;t)\dot{x}^{2}+\dot{y}^{2}))) & = & 0,
\end{eqnarray*}
\begin{eqnarray*}
\frac{d}{dt}(\frac{\text{\ensuremath{\partial}}}{\text{\ensuremath{\partial}}\dot{y}}(\frac{1}{2}(a_{xx}(x,y;t)\dot{x}^{2}+\dot{y}^{2})))-\\
\frac{\text{\ensuremath{\partial}}}{\text{\ensuremath{\partial}}y}(\frac{1}{2}(a_{xx}(x,y;t)\dot{x}^{2}+\dot{y}^{2})) & = & 0.
\end{eqnarray*}
This allows us to find the equation of motion for the y component
as
\[
\frac{d}{dt}\left(\dot{y}\right)=0.
\]
Take into account that the $p_{y}=\dot{y}$, we find that the y component
of the momentum is conserved. During the particle reflection from
each wall of the square boundary, this component of the momentum would
be the normal or the tangent component of the momentum and its value
remains unchanged. Therefore, its conservation is preserved in the
square billiard.

Moreover, considering $\omega=0$ in the static situation, it is easy
to show that the $p_{y}$ is conserved in the static case, too. In
addition, for the static case the energy is also conserved. So, the
billiard in the static situation has two constants of motion. The
phase plot of a static two dimensional billiard has a 4 dimensional
phase space. When there are two constant of motion, the particle trajectory
would be placed in a two dimensional subspace of the phase space.
Therefore, the system which corresponds to the Fig. \ref{fig. 10},
is integrable. So, as mentioned before, the direction of the wave
fronts with respect to the boundary plays an important role in preserving
symmetries of the system.

The integrable dynamics implies no energy growth for the bouncing
particle.\textcolor{cyan}{{} }

We expect that if the static version of this non-planar time-dependent
billiard is integrable, then its energy wouldn't grow. We also expect
from the LRA conjecture that, when, the freezed system is chaotic,
the particle energy grows unbounded. But here the particle energy
for large number of iterations converges to a constant value. Thus,
our results don't agree with this conjecture for large number of collisions.

\section{Conclusion}

We studied the dynamics of a particle in a non-planar billiard under
two different conditions: (i) when the surface of the billiard is
non-planar and static. (ii) when the surface of the billiard is non-planar
and changes with time. We assumed that, the plane of the billiard
is a sinusoidal traveling wave surface. 

We presented a numerical investigation of a time-dependent non-planar
billiard. We explained the behavior of trajectories, the phase space
and the average energy growth. We showed that a trajectory becomes
irregular by increasing either of wave vector, $k$, or wave amplitude,
$A$. We displayed the reduced phase space plot of the static system
for some cases. In addition, we explained the influence of different
control parameters of the non-planar surface on the dynamics of the
static system. We showed that, as the billiard gets non-planar, the
dynamics of the billiard transits from regular to mixed and then it
gets completely chaotic. Finally, we showed that depending on the
parameters of the system the particle energy may exhibits a limited
growth\textcolor{red}{{} }or stays near a constant value. Considering
the dynamics of their static case, the results don't exhibit unbounded
energy growth which is expected from the LRA conjecture.\textcolor{red}{{}
}As explained before, the LRA conjecture states that ``chaotic dynamics
of a billiard with a fixed boundary is a sufficient condition for
the Fermi acceleration in the system when a boundary perturbation
is introduced'' \cite{Loskutov2000}. We studied the behavior of
the average energy growth in the non-planar time-dependent billiard.
We showed that when in the static case, the billiard is chaotic, then
the particle energy in the time-dependent billiard doesn't grow unbounded.
Therefore, the LRA conjecture doesn't necessarily hold in this condition.
We also emphasize that our results doesn't contradict the LRA conjecture,
since, the LRA conjecture is proposed for billiards with time-dependent
boundary. Our results state that this conjecture doesn't extend to
the time-dependent non-planar billiard model. 

For future work, a relativistic treatment of this model is interesting. 

We also believe that the sensitive dependency of the dynamics of the
system on the features of traveling wave (like its amplitude, wave
length and direction of its entrance into the billiard) could be used
for sensing applications. For example it can be used to detect small
perturbations in the geometry of a flat plane and even to measure
their characteristics.

We can simulate the behavior of a particle in a billiard in a space-time
with a time-dependent metric of general relativity like gravitational
waves. The results can be used for detecting gravitational waves. 

\appendix

\section*{Appendix}

In this Appendix, we explain the derivation of Euler-Lagrange equation
of motion. Lagrangian of a particle constrained to a curved surface
has the form
\[
L=\frac{1}{2}a_{ij}\dot{u}^{i}\dot{u}^{j},
\]
where the $a_{ij}$'s in general, are functions of u's and possibly
also of the time.

The Euler-Lagrangian equation can be obtained using the following
equation
\[
\frac{d}{dt}\left(\frac{\partial L}{\partial\dot{u}^{i}}\right)-\frac{\partial L}{\partial u^{i}}=0.
\]
This gives
\begin{eqnarray*}
\frac{d}{dt}\left(\frac{\partial L}{\partial\dot{u}^{i}}\right) & = & \frac{d}{dt}\left(\frac{\partial\left(\frac{a_{mk}\dot{u}^{m}\dot{u}^{k}}{2}\right)}{\partial\dot{u}^{i}}\right)=\frac{d}{dt}\left(a_{ik}\dot{u}^{k}\right)\\
\, & = & a_{ik}\ddot{u}^{k}+\frac{\partial a_{ik}}{\partial u^{m}}\dot{u}^{m}\dot{u}^{k}+\left(\frac{d}{dt}a_{ik}\right)\dot{u}^{k}.
\end{eqnarray*}
\begin{eqnarray*}
\frac{\partial L}{\partial u^{i}} & = & \frac{\partial\left(\frac{a_{mk}\dot{u}^{m}\dot{u}^{k}}{2}\right)}{\partial u^{i}}\\
\, & = & \frac{1}{2}\frac{\partial a_{mk}}{\partial u^{i}}\dot{u}^{m}\dot{u}^{k}.
\end{eqnarray*}
Hence the Lagrange-Euler equations of motion reads
\begin{eqnarray*}
\frac{d}{dt}\left(\frac{\partial L}{\partial\dot{u}^{i}}\right) & = & a_{ik}\ddot{u}^{k}+\frac{\partial a_{ik}}{\partial u^{m}}\dot{u}^{m}\dot{u}^{k}+\left(\frac{d}{dt}a_{ik}\right)\dot{u}^{k}\\
\, & = & \frac{\partial L}{\partial u^{i}}=\frac{1}{2}\frac{\partial a_{mk}}{\partial u^{i}}\dot{u}^{m}\dot{u}^{k},
\end{eqnarray*}
\[
a_{ik}\ddot{u}^{k}+\partial_{m}a_{ik}\dot{u}^{m}\dot{u}^{k}+\left(\frac{d}{dt}a_{ik}\right)\dot{u}^{k}=\frac{1}{2}\partial_{i}a_{mk}\dot{u}^{m}\dot{u}^{k}.
\]
Notice that
\[
\partial_{m}a_{ik}\dot{u}^{m}\dot{u}^{k}=\frac{1}{2}\left(\partial_{m}a_{ik}\dot{u}^{m}\dot{u}^{k}+\partial_{k}a_{im}\dot{u}^{m}\dot{u}^{k}\right).
\]
Thus we get to the equation bellow
\[
a_{ik}\frac{d^{2}u^{k}}{dt^{2}}+\frac{1}{2}\left(\partial_{m}a_{ik}+\partial_{k}a_{im}-\partial_{i}a_{mk}\right)\dot{u}^{m}\dot{u}^{k}+\left(\frac{d}{dt}a_{ik}\right)\dot{u}^{k}.
\]
\[
\frac{d^{2}u^{i}}{dt^{2}}+\frac{du^{m}}{dt}\left[mk,i\right]\frac{du^{k}}{dt}+\left(\frac{d}{dt}a_{ik}\right)\dot{u}^{k}=0.
\]
Where, $\left[mk,i\right]=\frac{1}{2}\left(\partial_{m}a_{ik}+\partial_{k}a_{im}-\partial_{i}a_{mk}\right)$.

This is the Euler-Lagrange equation of motion of a free particle on
a curved time-dependent surface.

For more details see \cite{dewitt1957dynamical}. 

\bibliographystyle{apsrev4-1}
\addcontentsline{toc}{section}{\refname}\bibliography{new_ref}

\end{document}